# A Framework for Information Disorder: Modeling Mechanisms and Implications Based on a Systematic Literature Review

*Julie* Ricard[a], *Ivette* Yañez[b], *Leticia* Hora[c].

[a] Fundação Getulio Vargas / CNPq, São Paulo, Brazil, julie.ricard@fgv.edu.br, ORCID 0000-0003-2304-5781.

[b] Data-Pop Alliance, Mexico City, Mexico, iyanez@datapopalliance.org, ORCID 0009-0007-9415-5490.

[c] Universidade de São Paulo, São Paulo, Brazil, leticiahora1@usp.br, ORCID 0000-0001-5277-0312.

**Abstract**

This systematic literature review seeks to explain the mechanisms and implications of information disorder for public policy and the democratic process, by proposing a five-stage framework capturing its full life cycle. To our knowledge, no prior reviews in the field of public administration have offered a comprehensive, integrated model of information disorder; most existing studies are situated within communication, information science, or data science, and tend to focus on isolated aspects of the phenomenon. By connecting concepts and stages with enabling factors, agents, tactics and impacts, we reframe information disorder not as a question of "truthiness," individual cognition, digital literacy, or merely of technology, but as a socio-material phenomenon, deeply embedded in and shaped by the material conditions of contemporary digital society. This approach calls for a shift away from fragmented interventions toward more holistic, system-level policy responses.

**Introduction**

Since 2016, disinformation has emerged as a critical public concern, prompting a surge in academic research (Wang et al., 2022). High-profile cases such as the 2016 U.S. elections and Brexit have demonstrated its power to sway public opinion (Berlinski et al., 2023), disrupt democratic processes (McKay & Tenove, 2021), deepen polarization (Marlow et al., 2021), and threaten public health, exemplified by misinformation during COVID-19 (Dowling & Legrand, 2023).

As reflected in the centrality of social media in studies in the field, disinformation is overwhelmingly associated with the rise of platforms. Indeed, most existing literature reviews (Machete & Turpin, 2020; Celliers & Hattingh, 2020; Aïmeur et al., 2023) focus exclusively on disinformation within social media platforms. There is a pronounced emphasis on high scale detection of mis- and disinformation (Nguema et al., 2024; Machete & Turpin, 2020), particularly from a highly technical standpoint, with a focus on machine learning techniques for identification. In that sense, research often isolates the phenomenon from its broader socio-political contexts, largely neglecting traditional media and offline dynamics. There remains a need for holistic frameworks that capture disinformation's full life cycle—from enabling sociopolitical and informational ecosystem vulnerabilities, to the motivations of its agents, the tactics they employ, and the resulting impacts on both the information environment and society at large.

In this article, we adopt the term "information disorder" to emphasize our systemic approach to the phenomenon, spanning production, circulation, and reception of false, misleading or distorted information (Posetti et al., 2018; Baron, 2024). We shift away from seeing disinformation as solely a matter of information and technology. Instead, we consider information disorder as a relational phenomenon deeply embedded in contemporary digital and social infrastructures, shaped by vulnerabilities, institutional dynamics, ideological projects, and informational ecosystems. Under this umbrella, we include a range of terms used across disciplines—such as misinformation, propaganda, conspiracy theories, and fake news—recognizing their overlapping functions in shaping public discourse, and influencing democratic governance and policy processes.

To address existing gaps, we propose a systematic literature review structure in two phases. The first phase maps how disinformation is studied within public administration, political science, and international relations, identifying existing themes and research gaps (Ricard et al., 2024). The second, core phase proposes a theoretical framework based on the following guiding research questions:
1. What sociopolitical and informational factors enable information disorder?
2. Who are the main actors involved, and what are their motivations?
3. What tactics do these actors use to fuel information disorder?
4. How does information disorder transform the broader information environment?
5. What is its impact on society, democratic institutions, and public policy?

Notably, we found no systematic literature reviews on mis- and disinformation—or the broader concept of information disorder—within the fields of public administration or political science. To help fill this gap, we adopt a model-oriented approach to propose an integrative theoretical framework that maps the interrelationships among the core components of information disorder. This includes identifying the feedback loops that sustain it over time. In doing so, the framework aims to support more coordinated, system-level responses across both research and policy domains.

## Methodology for Systematic Literature Review

Our systematic literature review was conducted in several stages, following established protocols for transparency and replicability. The process began with the selection of keywords, followed by database searches, filtering and screening of articles through a double-blind review process, and finally, a coding and analysis phase.

### Choice of Keywords

The selection of keywords for this systematic literature review was developed through an iterative process among co-authors. Based on the goal and key questions driving this review, we listed the core constructs of our research (column 1 of table below). It is informed by established literature typologies on disinformation and the typical phases observed in disinformation literature: causation/creation, propagation/spread, and consequences/impacts.

| Core constructs | Keywords |
| --- | --- |
| information disorder | ("information disorder" OR "misinformation" OR "disinformation" OR "fake news" OR "conspirac*" OR "manipulation" OR "information war*" OR "deep fak*" OR "infodemi*" ) AND |
| AND | |
| public policy | ("government policy" OR "public polic*" OR "policy mak*" OR "policy-mak*" OR "public administration" OR "public sector" OR |
| democratic process | "democratic process" OR "democratic institutions" OR "electoral system" OR "election*" OR "voting" OR "civi* engagement" OR "activism" OR "democratic" OR |

|  |  | "democracy" OR "public debate" OR "public sphere") AND |
| --- | --- | --- |
| AND |  |  |
| creation/ prod | WHY | ( "root*" OR "cause*" OR "origin" OR "emergence" OR "creation" OR "production" OR "risk factor*" OR |
| distribution | HOW | "seed*" OR "spread*" OR "proliferation" OR "propagation" OR "distribution" OR |
| consequences | EFFECTS | "impact*" OR "consequence*" OR "influence" OR "effect*" OR "outcome*") |

*Table 1: Keyword choice for the Literature Review. Source: elaborated by the authors.*

**Search in Databases**

The keywords above were used to conduct searches in four databases: Web of Science, Scopus, SciELO and Association for Information Systems (AIS). Web of Science and Scopus are the most often used databases for literature reviews, given their extensive coverage and authoritative presence in academia, providing a robust foundation of peer-reviewed sources. To encompass research from Latin America, we also conducted the search in SciELO, and translated all keywords to Spanish and Portuguese to take advantage of the trilingual aspects of this repository. Additionally, the AIS database was included to capture the vanguard research in the field of information systems, thus ensuring that our review encompasses cutting-edge insights into the technological aspects of disinformation.

We conducted the keyword searches in titles, abstracts and author keywords. In Web of Science and Scopus, we filtered the results using subject/research areas, as outlined below, in order to limit the results to the fields of Public Administration, Political Science and International Relations. We also filtered by years (2016-2024), considering the boom in disinformation research since the 2016 USA elections.

| Web of Sciences | | |
| --- | --- | --- |
| Category | Social Sciences | |
| Research Areas | Public Administration | |
|  | Political Science | |
|  | International Relations | |
| **Scopus** | | Code |
| Subject Area | Social Sciences & Humanities | |
| ASJC category | Political Science and International Relations | 3320 |
|  | Public Administration | 3321 |

*Table 2: Search by category in Web of Science and Scopus. Source: elaborated by the authors*

**Double-blind Screening Process**

Complying with systematic literature review protocols, we rigorously applied the double-blind review in the screening process. Initially, we extracted and merged articles from four distinct databases, with the Web of Science structure serving as the base for this merge. This step was followed by the removal of duplicates – through which we reduced the number of articles to 840. Ahead of the following steps, we introduced a coding system according to criteria for inclusion, where A=Definitely include, B=Maybe include, and C=Definitely do not include. In the **title screening** phase, only articles that did not explicitly relate to disinformation (marked C-C) were excluded. We then advanced to the abstract screening phase, which was conducted in two stages. Our guiding inclusion criteria was to retain articles that provide insights into typical phases of disinformation (emergence, propagation, effects). We excluded studies with a focus on countermeasures or those primarily examining individual-level risk factors, as well as book reviews. With the above-mentioned criteria in mind, we discussed amongst co-authors the articles marked A-B, A-C and B-B, bringing our total to 195 articles. We then proceeded to a filter by journal quality, keeping only articles from Q1 Journals as per SJR classification. Articles pertaining to a journal ranked Q2, Q3, or Q4, to a journal not ranked as per SJR classification or to a conference, were only included if their geographic focus was a country of the Global South. We did so in order to ensure a minimum level of diversity in the sample of articles, considering that the production in the field is overwhelmingly focused on the United States (Wang et al., 2022). In the case of conference proceedings, we included articles more recent than 2021, to ensure we capture the latest advancements in terms of research. Finally, we included additional 4 articles from the field of public administration found through references. Altogether, this brought the total number of articles to be assessed for eligibility to 121 articles (Figure 1). **After we did a full article review, a final total of 95 articles were selected for this review.**

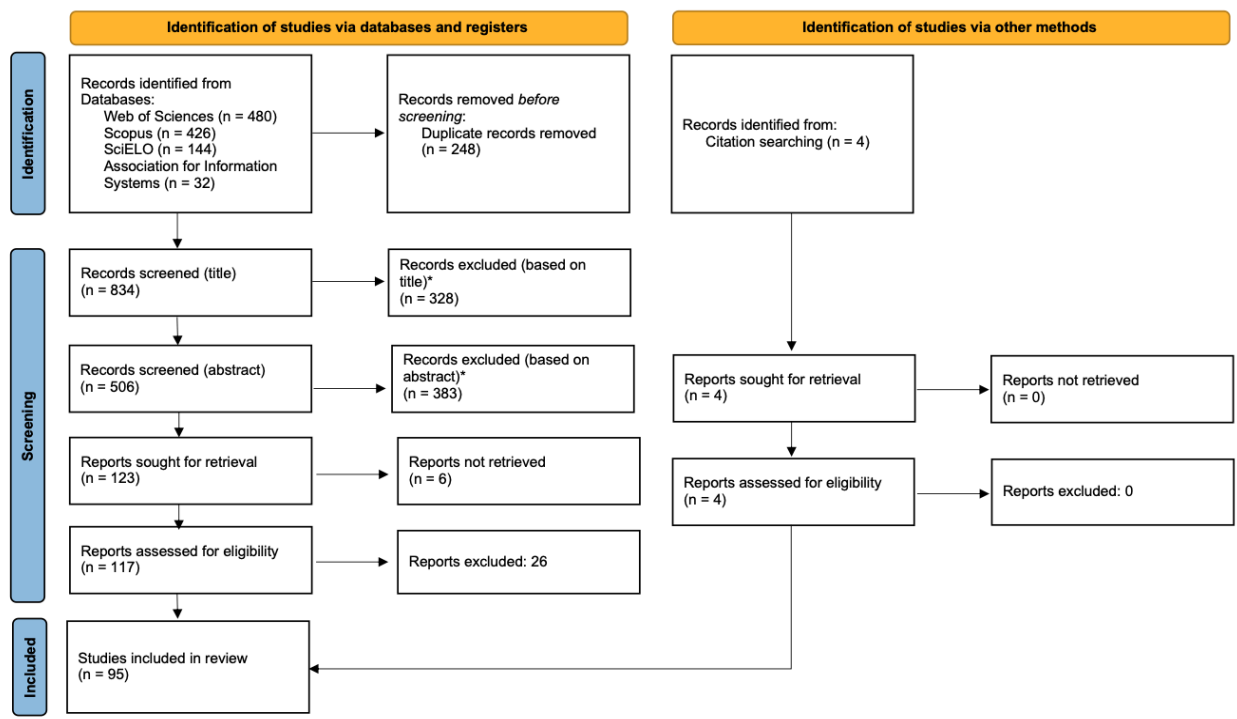

Figure 1: PRISMA Flow Diagram for the Literature Review. Source: Elaborated by the authors, based on the template made available by Page MJ, et al. BMJ 2021;372:n71. doi: 10.1136/bmj.n71, under CC BY 4.0.

**Codification and Analysis**

To guide our codification process, we drew inspiration from the CIMO logic (Denyer & Tranfield, 2009), a heuristic commonly used in systematic reviews to map the interplay between Context, Intervention, Mechanism, and Outcome. This framework helped inform the structure and focus of our codes, enabling us to capture not only who is involved in disinformation (agents), but also where and how it unfolds (context and mechanisms), and with what consequences (outcomes). While our codes were refined iteratively during the reading and discussion of articles, they were grounded in this analytical logic to support causal and explanatory insights—rather than purely descriptive classification.

To streamline the coding process and facilitate the subsequent analysis, we used MAXQDA for coding the articles, then proceeded with merging and exporting the codified segments into a csv. The final codes (15 in total) as well as the number of segments codified for each are further described in Table 3. As part of the analysis, co-authors held biweekly calls to discuss insights from the analysis, throughout the codification period (4 weeks).

| Context | Mechanisms |
|---|---|
| <ul><li>Country/countries or region(s) primarily analyzed in the study. *Number of segments codified: 93*</li><li>Field of Knowledge/Theoretical Frameworks. *Number of segments codified: 366*</li><li>Political Science vs. Public Policy Angles? *Number of segments codified: 145*</li><li>Methodological Approach. *Number of segments codified: 380*</li><li>Definitions of Disinformation. *Number of segments codified: 183*</li><li>Vulnerability Factors. *Number of segments codified: 231*</li><li>Context and Roots: Why the Narrative Works. *Number of segments codified: 324*</li></ul> | <ul><li>Reasons Behind Disinformation ("raison d'être"). *Number of segments codified: 143*</li><li>Strategies of Legitimation. *Number of segments codified: 81*</li><li>Specific Topics of Disinformation Addressed. *Number of segments codified: 187*</li></ul> |
| **Interventions** | **Outcomes** |
| <ul><li>Stakeholders Involved in the Information Ecosystem and How. *Number of segments codified: 251*</li><li>Tactics for Spreading Disinformation. *Number of segments codified: 459*</li><li>Type of Media or App Studied. *Number of segments codified: 177*</li></ul> | <ul><li>Impact of Disinformation. *Number of segments codified: 371*</li><li>Solutions. *Number of segments codified: 89*</li></ul> |

*Table 3: Codification for analysis. Source: elaborated by the authors.*

**Results: Towards a Framework for Information Disorder**

As mentioned previously, we systematically analyzed findings from 95 academic articles to construct a comprehensive model that explains the mechanisms and implications of information disorder for both democracy and policy-making. The development of this framework was guided by a multidisciplinary theoretical approach, incorporating perspectives from various fields to provide a robust understanding of the topic, and by grouping the outcomes from the 15 codes listed in the previous section into coherent parts of our proposed framework.

The proposed model is structured around five key stages and feeding loops:

1. **Enabling Factors and Contextual Vulnerabilities** – This stage examines the sociopolitical, and informational factors that create an environment susceptible to information disorder in a given setting.
2. **Exploitation by Motivated Disinformation Agents** – This stage describes how these enabling vulnerabilities are exploited or leveraged by agents of information disorder, who usually hold institutional power and operate with specific objectives standing to benefit from economic, social or political gains.
3. **Tactics Leveraged at Different Levels** – Here, we explore the strategies and tactics used by these agents, ranging from message crafting and platform exploitation, to information flow manipulation and media infiltration.
4. **Impact on the Information Environment** – This stage analyzes how information disorder directly affects the broader information ecosystem.

5. **Impacts on Society, Policy and Democracy at Large** – Finally, we consider the real-world impacts of these disrupted information environments on society -particularly in relation to public policy and democracy, but also including further entrenching vulnerabilities (see stage 1).

The model suggests a cyclical process where pre-existing vulnerabilities are exploited by disinformation agents, who then propagate disinformative narratives using various tactics. These narratives directly influence the information ecosystem, leading to tangible societal impacts, which in turn may reinforce the original vulnerabilities in any given context. In the sections that follow, we detail the mechanisms operating within and between each stage, illustrating how they interact to sustain and perpetuate information disorder and its effects on democratic processes and policy outcomes.

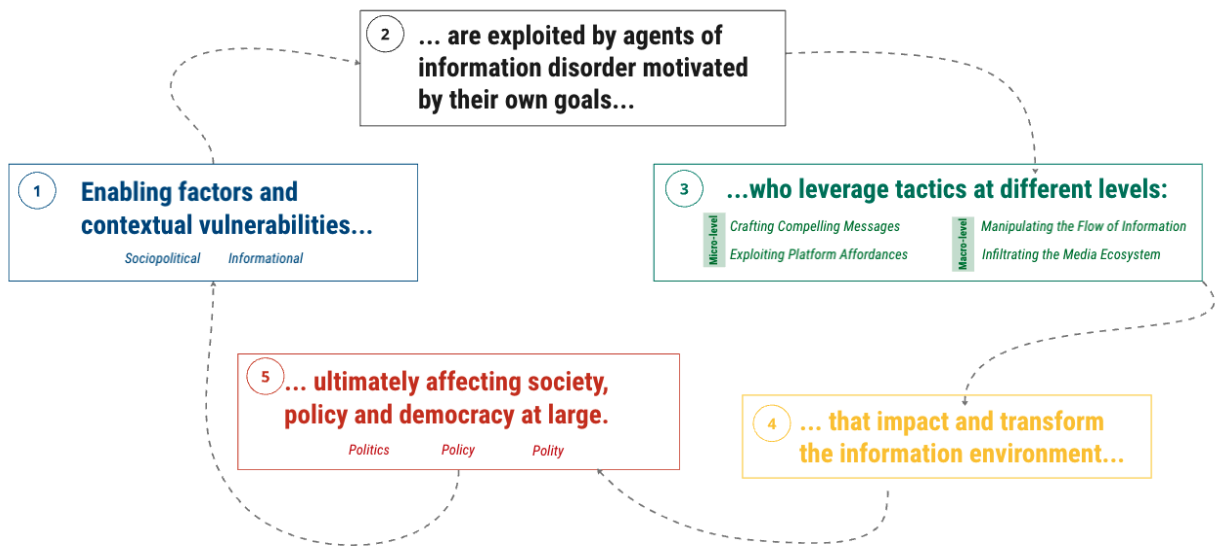

*Figure 2: Proposed Framework for Information Disorder. Source: elaborated by the authors.*

### 1. Enabling Factors and Contextual Vulnerabilities

Our analysis revealed a wide range of enabling factors that have been exploited by agents of information disorder in different contexts. These "contextual vulnerabilities" were identified through the coding of articles under the categories "Context and Roots: Why the Narrative Works" (Number of segments codified: 324) and "Vulnerability Factors" (Number of segments codified: 231). Based on these findings, we categorized the factors into two main types: sociopolitical vulnerabilities, and informational vulnerabilities. Sociopolitical and informational vulnerabilities often overlap creating specific intersections of risk. Within each category, specific factors can be singled out to different extents in different contexts, and have more or less relevance accordingly. More research is needed to understand how certain vulnerability factors interact and can create particularly dangerous situations.

It is worth mentioning that, despite limiting the scope of this research to the fields of political science, public administration and international relations, we obtained articles at the intersection of one of these fields and other relevant fields, particularly psychology. As such, certain studies focus on cognitive biases and psychological traits that may contribute to an individual's susceptibility to disinformation. Jiang (2023) and Enders & Smallpage (2019) argue that "motivated reasoning" plays a critical role in how individuals process information, often aligning it with their desires and preconceptions, particularly in politically polarized environments. A deterministic mindset, as noted by Jiang (2023), fosters a fixed belief that events are predestined, making it difficult for individuals to accept new or contradictory information. Similarly, the "authoritarian personality", explored by Yendell & Herbert (2022), predisposes individuals to adhere to established hierarchies and resist information that challenges their worldview.

Overconfidence, highlighted by Partheymüller et al. (2022), further exacerbates this issue, as individuals with unwarranted confidence in their knowledge are more likely to believe and disseminate misinformation. Additionally, confirmation bias (Heckler & Ronquillo, 2019) and conspiratorial thinking (Yair et al, 2024; Martini et al., 2022) reinforce these tendencies, making individuals more resistant to corrective information. Once misinformation is embedded in a person's memory, cognitive entrenchment occurs, making retractions ineffective and often reinforcing the initial misinformation, creating a persistent vulnerability. Moreover, (Pickel et al., 2022) highlights the elective affinity between superstition, esotericism, and a conspiracy mentality.

Finally, certain studies mention demographic variables such as age (Yendell and Herbert, 2022; Nguyen, 2023; Cortina & Rottinghaus, 2022), gender, education level, and socioeconomic status as factors that may help explain why certain groups are more vulnerable to disinformation than others. For example, Nguyen (2023) finds "intergenerational divides in political information seeking", and Cortina & Rottinghaus (2022) that age seems to have an effect on belief of conspiracy theories. Yendell and Herbert (2022) used a survey on right-wing extremism and racism in the UK, controlled by age, gender and education level. They conclude that while gender does not play a role, age and low education are weakly correlated with conspiracy mentality. However, these basic sociodemographic variables (age, gender, education level) are mostly used as control variables, and often without very conclusive evidence.

These vulnerabilities at the individual level have purposefully not been included in the framework, for two main reasons. First, according to the scope of this research, we have only obtained a limited number of studies in this field, and therefore are not able to provide a comprehensive analysis of such results. Second, the goal of our research is to move beyond potentially deterministic factors at the individual level, towards an understanding at the collective/social levels of the mechanisms of information disorder. Within the literature itself, there is an underlying debate as to whether individual factors matter more than contextual factors. For example, Yendell & Herbert (2022) consider that "to explain generic conspiracy thinking we must look at people's characteristics more than the characteristics of the country in which they reside". On the other hand, Czech (2022) highlights that "in the case of attitudes, context matters more than personality traits or cognitive styles". Further research focused specifically on this debate would be needed to provide more conclusive evidence. Drawing on Latour's actor-network theory (2005), we argue that knowledge and social phenomena emerge from assemblages of human and non-human actors (e.g. technologies, devices, media platforms), thus undermining human-centric explanations that focus only on individual minds. As such, information disorder is best explained by the socio-technical networks and environments that produce and propagate mis-disinformation, rather than by psychological characteristics or deficit of their audience.

Ahead, we further explain the typology created and provide a detailed description of each enabling factor.

1. **Sociopolitical Vulnerabilities**: This category includes socioeconomic, political, and cultural contexts that disinformation agents can exploit to deepen existing divisions within a society. This was identified as the most significant category, with numerous studies demonstrating how historical grievances, political tensions, cultural norms, and societal cleavages are strategically leveraged, reaped and shaped into specific narratives to legitimize and anchor disinformation narratives. These vulnerabilities are particularly potent because they often touch on deeply ingrained societal issues, enabling disinformation to thrive by making it more resonant and harder to counter.
2. **Informational Vulnerabilities**: These refer to weaknesses in the information environment often resulting from technological shifts, such as new business models, media consumption habits and the decline of mainstream journalism, which can be weaponized by disinformation agents. With the fragmentation of information sources and the rise of social media and other digital platforms, the traditional gatekeeping role of journalism has weakened, creating an environment where disinformation can spread more easily and faster, reach larger audiences due to its inflaming narratives being fueled by the 'attention economy' that rules algorithms, and ultimately be perceived as credible.

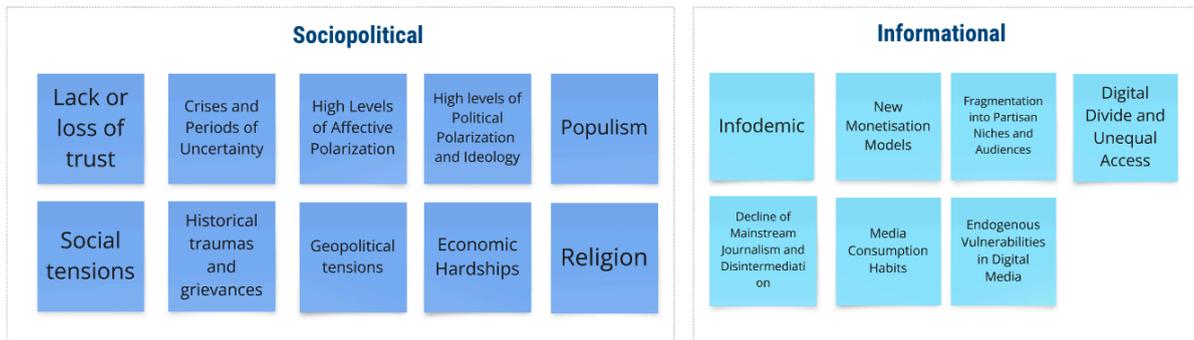

*Figure 3: Proposed Enabling Factors and Contextual Vulnerabilities. Source: elaborated by the authors.*

**Sociopolitical Vulnerabilities**:

- **Lack or loss of trust.** The erosion, loss and/or absence of trust in key institutions appears as a key socio political vulnerability when it comes to disinformation. A lack of trust in government, as discussed by Islam et al (2023), Dowling & Legrand (2023), and Zimmermann (2020), or in public institutions (Koc-Michalska et al., 2023), or in media (Freiling et al., 2023) – is a vulnerability factor that can be exploited at both the individual and sociopolitical levels. The systematic erosion of democratic norms and trust in institutions over time weakens societal resilience to disinformation, making it easier for such narratives to take hold. Note that this factor can be both a vulnerability and an impact of information disorder, creating a feedback loop.
- **Crises and periods of uncertainty**. Disinformation thrives in times of crisis and uncertainty, such as health crises and elections. Kweon (2023) and Pickel et al (2022) demonstrate how susceptibility to conspiracy myths increases during crises, such as the Covid-19 pandemic that acted as a significant catalyst for such narratives. Bradshaw (2023) also notes how heightened public anxiety can provide fertile ground for disinformation to take root. The author explores how Russia exploited the tensioned context of the Black Lives Matter (BLM) protests to share divergent news about the same content in an attempt to manipulate and fuel political divisions in the media, using networks such as Russia Today (RT) and Sputnik. Similarly, periods of heightened pre-election anxiety follow similar patterns, as discussed by Karekwaivanane (2019).
- **High Levels of Affective Polarization:** "Affective polarization" is an understanding of political polarization, explained by the strong identification of individuals with their political in-group and hostility towards out-groups. Jenke (2022), finds that affective polarization significantly increases susceptibility to misinformation that aligns with in-group beliefs. As discussed by Jenke (2022), this polarization is further intensified by political sophistication, suggesting that anyone, even well-educated individuals are not immune to disinformation when it resonates with their partisan identities. Ndahinda & Mugabe (2024) describe a similar dynamic of "us vs them narratives", promoted by hate speech used to fuel division and anti-Banyamulenge and anti-Tutsi violence in the Democratic Republic of Congo.
- **High levels of political polarization and ideology**: We find that polarization is often associated with disinformation (Faris et al., 2017 cited in Keller et al., 2020), (Jian 2023), (Lombana-Bermudez et al., 2022), (Jungherr & Rauchfleisch, 2024). For example, Mauk & Groping (2024) find that disinformation is a factor explaining polarization of beliefs about electoral integrity. Beyond electoral contexts, Gil de Zuniga et al. (2023) finds that different types of polarization – affective, ideological, and perceived societal– have a negative role on political persuasion through social media. Stabile et al. (2019), Enders et al. (2020), Kweon & Choi (2023), and Jiang (2023) emphasize that individuals with strong partisan or political identities are more prone to accepting and sharing misinformation. This susceptibility is found to be particularly pronounced among conservatives, as indicated by studies such as Guess, Nyhan, and Reifler (2020) cited by Hughes & Waismel-Manor (2021), and further supported by Kweon & Choi (2023), Enders & Smallpage (2019), Jungherr & Rauchfleisch (2024), and Yendell and Herbert (2022).

- **Populism**: Studies suggest that populism both favors and is favored by disinformation. Populist governments are pinpointed as particularly vulnerable contexts for disinformation to thrive (Arceneaux & Truex, 2023), (Massari 2018 cited by Lôbo & Morais-2019), to the point of finding that individuals in countries with populist governments were more likely to believe conspiracy theories (Kweon & Choi, 2023). The rhetoric of leaders like Trump, which builds on long-term trends such as delegitimizing political opponents, attacking the media, and appealing to racism and xenophobia, creates an environment ripe for disinformation (Edwards III, 2020). At the same time, research suggests that believing in disinformation leads to supporting populism. Serani (2023) finds that "individuals who have anti-vax attitudes and who also have a higher propensity to believe in conspiracy theories are more likely to vote for populist parties".

- **Social tensions**. Disinformation is often tailored to exploit societal vulnerabilities (Bjola & Papadakis, 2020), (Balfour, 2020) around "emotionally intense issues" (Bjola & Papadakis, 2020) —such as child-parent separation, same-sex relations, and minority rights—to provoke strong public reactions. Racial tensions tensions are the most frequently studied in our review (Yendell & Herbert, 2022), (Gunther at al., 2019), (Pickel et al, 2022), (Heckler & Ronquillo, 2019), (Vićić & Gartzke, 2024). For example, it is argued that Russian disinformation campaigns in the US have purposely targeted African Americans, exploiting systemic inequalities and historical disenfranchisement to fuel disinformation narratives (Vićić & Gartzke, 2024). Similarly, xenophobic tensions are weaponized in Europe, for example against immigrants (Bjola & Papadakis, 2020) and minority groups like Muslims (Stoeckel, 2023). In Africa, ethnic tensions are leveraged to mobilize fear and insecurity (Ndahinda & Mugabe, 2024), (Sahar & Sahar, 2021). Gender tensions and stereotypes are also found to be weaponized (Stabile et al, 2019).

- **Historical traumas and grievances.** Findings highlight how disinformation narratives can invoke historical traumas and grievances. This can happen within countries, for example in the case of Vietnamese Americans (Nguyen et al., 2023), but it can also tap into unresolved historical conflicts, such as the historical distrust between Russia and the West. These historical references can resonate powerfully with audiences, making the disinformation seem more plausible (Perl et al., 2018).

- **Geopolitical tensions**. Disinformation narratives are often anchored in ongoing political and geopolitical tensions, thus contributing to so-called "information warfare" (Zhao et al., 2023). For example, tensions between NATO, the U.S., Russia, and Ukraine are leveraged to anchor narratives that favor specific geopolitical agendas. Examples include propaganda disseminated by Russia to influence EU trade deals with Ukraine, or narratives about combating "terrorists" to justify the intervention in Syria –along the lines of what U.S. portraits–, which was later debunked by NATO (Bolton 2021). Studies also point to campaigns with foreign interference purposes, such as social media disinformation campaigns conducted by Russia's Internet Research Agency (IRA) in the US, especially focusing on African Americans (Vićić & Gartzke, 2024) and influence operations in the European Union (Dawson & Innes, 2019). Within countries, political divides can also be exploited by foreign actors, for example to erode trust in Western democratic processes (Justwan & Williamson, 2022).

- **Economic Hardships**. Economic concerns, such as fears over job security or the impact of sanctions, are often weaponized in disinformation campaigns or simply contribute to anchoring information disorder narratives. Kermani (2023) and Hughes and Waismel-Manor (2021) highlight how economic hardships can be exploited to fuel disinformation, tapping into the public's economic insecurities.

- **Religion.** Religiosity is mentioned in multiple articles (Czech, 2022; Hidalgo, 2022; Yendell & Herbert, 2022) as a potential factor of social vulnerability to disinformation. Interestingly, it is not individual spirituality that seems to matter, but rather the type of religion Pickel at al (2022). More specifically, religions with authoritarian qualities are associated with more tendency towards conspiracy beliefs, such as religious fundamentalism (Yendell & Herbert, 2022) or Polish Catholic nationalism (Czech, 2022). It is argued that religion and conspiracy theories can have similar operational systems, and that in some settings, religion becomes "a socially rooted set of beliefs integrated within the political program" (Czech, 2022). Hidalgo (2022) agrees when he states that "both religions and conspiracy theories work as moral, intellectual, and political authorities within modern societies, at least for the believers of corresponding religious or conspiracy- theoretical messages. In this respect, they each become an entity that normally does not lose its persuasiveness through scientific research, empirical studies, or fact-checking; on the contrary, it is located in a vacuum in which (many) people continue to form their emotional identity and stability" (Hidalgo, 2022).

**Informational Ecosystem Vulnerabilities**:

- **Infodemic.** The relationship between media consumption habits and vulnerability to disinformation is complex. The sheer volume of information available online, often referred to as an "infodemic," poses a significant challenge. As Maati et al. (2023) points out, the massive quantities of information can overwhelm individuals, making it difficult to distinguish between accurate and misleading content. Make sure it's the right term and if it's actually mentioned by multiple authors in our review or not, also double check the definition from WHO to make sure it's not only related to health.
- **Attention Economy**. In the digital era, information disorder is magnified by the dynamics of the "attention economy," which incentivizes keeping users engaged with content for as long as possible. The algorithms acting as editors of the content that actually makes it to users' eyeballs, tend to favor compelling messages as described above. As Torreblanca (2023) notes, "negative or confrontational messages that deepen polarization" are particularly effective at generating user engagement and traffic on digital platforms. Bolton (2021) notes that "the digitalization of society has increased the potential to influence and manipulate cognition and emotions," likely because emotionally charged content tends to generate higher engagement on digital platforms (Torreblanca, 2023) - are more broadly discussed in tactics.
- **New Monetisation Models**: Driven by the logic of the attention economy—which incentivizes maximizing user engagement and traffic through sensational or polarizing content (Wu, 2016)—platforms' automated advertising systems not only permit but actively encourage the creation of websites that resemble legitimate media outlets (Torreblanca, 2023). These sites blend authentic news stories with misleading content to attract audiences and capture advertising revenue.
- **Decline of Mainstream Journalism and Disintermediation.** The decline of traditional journalism as a reliable arbiter of truth has significantly contributed to the rise of disinformation. According to Dahlgren (2018, as cited by Bolton, 2021), mainstream journalism has lost its role as the gatekeeper of truth due to the revolution in broadcast technology and the rise of social media. This disintermediation, as Torreblanca (2023) notes, has displaced traditional gatekeepers such as producers, editors, and scholars. While this shift has democratized the dissemination of information, it has also facilitated the widespread propagation of unvetted and potentially harmful content (Freiling et al. - 2023).
- **Media Consumption Habits**. The relationship between media consumption habits and vulnerability to disinformation is complex and not fully agreed upon in the literature. Some studies suggest that consuming news via social media increases vulnerability to disinformation due to the platform's structure and the prevalence of misleading information (Bjola and Papadakis, 2020; De Zúñiga, 2023; Keller, 2020). On the other hand, Gadjanova et al. (2022) discusses how even those who do not use social media are vulnerable, as they often receive second-hand information through friends and family, particularly in rural or socioeconomically disadvantaged communities. This "indirect user" phenomenon highlights that both users and non-users of social media are at risk, albeit in different ways.
- **Fragmentation into Partisan Niches and Audiences**. The information ecosystem has become increasingly fragmented, with audiences now divided into partisan niches that reinforce their pre-existing beliefs. Mauk & Grömping (2024) highlight how this fragmentation has led to the creation of echo chambers (Nonnecke et al., 2022), where individuals are only exposed to information that aligns with their views. Lombana (2022) and Krafft and Donovan (2020) further discuss the role of echo chambers in social media, which may amplify misinformation and disinformation within these closed groups. It is noteworthy that emerging research also questions the importance of echo chambers, arguing they may not be as prevalent as commonly accepted nor necessarily have determining effects on susceptibility to believing disinformation.
- **Endogenous Vulnerabilities of Digital Media.** The digital media landscape introduces new vulnerabilities that can be exploited for disinformation purposes. As Torreblanca (2023) points out, digital media allows for the manipulation of public opinion more quickly and effectively than traditional media. Endogenous vulnerabilities include:
    - **Mediation by Profit-Driven Algorithms:** The algorithms that govern social media platforms play a crucial role in what users see and how often they see it. Algorithms personalize and customize information consumption, making individuals more susceptible to tailored disinformation.
    - **Information Flow Instantaneity**: The rapidity and scale of information transmission in cyberspace, as noted by Vićić & Gartzke (2023), create an environment where disinformation can spread rapidly. Chernobrov (2022) identifies key dimensions of these vulnerabilities, including the immediacy and intimacy of news events, individuation, and interactivity. Scholars argue that the expansion of information disorder narratives has "stemmed from the suppression of physical barriers to access information, …and the feeling of direct cooperation with the political representatives" (Lôbo & Morais, 2021; Torreblanca, 2023).
    - **Opacity:** Even though external researchers have certainly found that algorithms prioritize, we do not fully understand exactly how these algorithms decide what takes precedence and what users see first. In addition, the use of artificial intelligence techniques hidden from plain sight are difficult to recognise, track, measure, and evaluate (Haleva-Amir, 2022). These automated decisions are opaque. In consequence, users who want to appear in prominent positions in platforms have incentives to try to force the ranking system that companies use (the algorithms) in their favour. Most of the time this is even done legitimately, i.e., by paying companies to achieve a better ranking through ads, for example. Others attempt the same by artificially and fraudulently generating traffic (Torreblanca, 2023).
    - **Lack of adequate filters and controls**: Another factor inherent to the digital ecosystem is the lack of adequate filters and controls, as networks are designed to facilitate access, not restrict it, making it easier for false information to pass as legitimate (Torreblanca, 2023).
- **Digital Divide and Unequal Access:** In addition to the stakeholders actions themselves, the research could also find narratives in which they draw attention to how other aspects related to internet use and access can impact the user's relationship with the content they find in the online environment. One notable example, is about the digital divide in Ghana media ecosystem, a complex system that impacts what Gadjanova et al. (2022) calls first-hand and indirect users, in which many Ghanaians are placed in a context where due to the difficult access to the internet impact indirectly to fake news exposure, in different levels based in their proximity with the use and access to internet.

**2. Exploitation by Motivated Disinformation Agents**

Scholars have called to investigate the variety of agents that contribute to information disorder –even if with varying levels of intentionality– (McKay & Tenove, 2021), as well as the motivations that drive them and the broader effects of their actions (Hunter, 2023). Our analysis revealed a wide range of both agents of information disorder –actors who deliberately produce, disseminate, or amplify misleading or false content within the information ecosystem– and underlying motivations. These agents operate with intention, often pursuing specific political, ideological, or economic interests. They include state actors, political figures, media influencers, grassroots groups, and ideological movements.

These elements were identified through the coding of articles under the categories "Stakeholders Involved in the Information Ecosystem and How" (number of segments codified: 251) and "Reasons Behind Disinformation ("raison d'être")" (number of segments codified: 143). Our analysis unfolds in two steps. First, we classify these agents according to the level at which they operate within the information disorder ecosystem, drawing from Bronfenbrenner's ecological model—ranging from interpersonal influence to cultural and institutional structures. Second, we analyze their underlying motivations or raison d'être—that is, the strategic objectives driving their disinformation practices, such as gaining political power, destabilizing opponents, deepening societal divisions, or seeking personal and financial gain (see Table 4). This two-layered approach enables a more nuanced understanding of how disinformation campaigns are not only about spreading falsehoods, but also about advancing broader agendas and reshaping public discourse.

Importantly, we do not seek to essentialize these actors or reduce them to fixed roles. Rather, it highlights the potential of these actors to contribute to information disorder depending on their context and interests. Many stakeholders—such as media organizations, governments, or grassroots groups—can either combat or propagate disinformation, depending on their position, incentives, and intent. This reinforces the need to analyze agents in conjunction with their motivations, or *raisons d'être.* Motivations also do not operate in isolation: different actors may share overlapping objectives or pursue multiple motivations simultaneously.

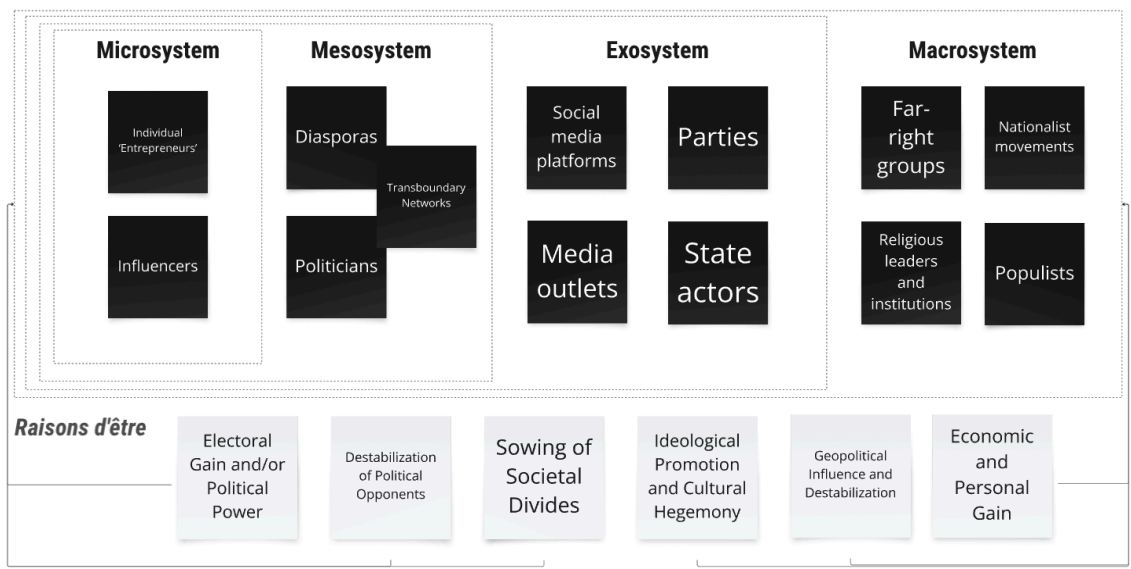

*Figure 4: Systematized table on Agents of Information Disorder and "Raisons d'être" Source: elaborated by the authors, based on the Ecological Model first proposed by Bronfenbrenner (1977).*

### 2.1 Agents of Information Disorder

The ecological model was first developed by Bronfenbrenner (1977) to explain human development. Over the years, the model has undergone several adaptations to reflect specific contexts and fields, ranging from **health behavior interventions (McLeroy et al., 1988)** to **violence prevention strategies (CDC,**

**2002)**. We leverage this model to classify agents according to the level at which they exert influence—ranging from interpersonal environments to cultural and ideological systems. Concretely, we classify the agents of information disorder are into four levels:

- **Macrosystem** – The macrosystem encompasses the overarching cultural, social, and ideological patterns that shape collective worldviews. This is where *agents promoting ideologies of culture* operate – such as political ideologues, nationalist movements, and religious institutions that articulate grand narratives and belief systems conducive to disinformation.
- **Exosystem** – The exosystem consists of institutions and infrastructures that shape the informational environment without direct interpersonal interaction. This is where *agents shaping the public agenda* operate – including the state, parties, media outlets and social media platforms, that indirectly condition the reach, repetition, and visibility of disinformation.
- **Mesosystem** – The mesosystem refers to networks and interactions across communities and local organizations. This is where *agents connecting and mobilizing communities* operate– such as diaspora networks and politicians, who relay and translate disinformation in culturally resonant ways.
- **Microsystem** – The microsystem involves the most immediate and personal layer in which individuals are embedded. This is where *agents exerting direct influence* – such as influencers and individual entrepreneurs — share and reinforce disinformation through trusted social bonds.

**Macrosystem: Agents Promoting Ideologies of Culture**

- **Far-right groups:** These groups exploit societal ideologies and divisions to spread extremist views and disinformation to influence cultural and political norms. (Keller et al., 2020)
- **Populist leaders:** Leaders or political figures who appeal to the masses by promoting simplistic, often polarizing, narratives that spread misinformation aligned with their populist rhetoric. Multiple scholars have highlighted Donald Trump's role in spreading misinformation and his impact on public discourse. Edwards III (2020) argues that Trump disseminated false information through various channels including social media, rallies, and television appearances. Balfour (2020) notes how Trump utilized institutional mechanisms through the White House to validate his narrative. (Casarões & Magalhães, 2021) further observes that in the United States and Brazil, "the alt-science network around the promotion of hydroxychloroquine came full circle - with presidents, politicians, media outlets, businesspeople, religious leaders, and scientists joining forces to advocate a fast and effective solution for the COVID-19 pandemic."
- **Nationalist movements:** These actors push disinformation that aligns with their ideological agenda of promoting national superiority or division. (Edwards III, 2020).
- **Religious leaders and institutions:** They shape cultural ideologies, sometimes promoting disinformation that aligns with their religious or cultural narratives. For example, in Gadjanova et al. (2022), authors explore how the role of religious leaders has been amplified with the use of social media, since their positions have an impact both online and physically in their communities.

**Exosystem: Agents Shaping the Public Agenda**

- **Media outlets:** Traditional media, including news outlets, play a large role in shaping narratives, either amplifying or countering disinformation. That being said, recent research has highlighted the diverse and complex role of the media in shaping the public agenda. Grossman (2022) suggests that viewing "media as an actor, with its own rules and dynamics, allows for a more nuanced perspective of media influence." This is particularly relevant when analyzing instances in which the media operates as an agent of information disorder, such as the case of the Russian state-controlled media ecosystem, which Bradshaw et al. (2023) describes as a coordinated union of various channels.
- **Social media platforms:** Platforms where disinformation spreads quickly due to algorithms that prioritize engagement, regardless of the content's veracity. One example is about the action of radical anti-democratic users that uses less regulated alternative social media (e.g. BitChute, Gab, Minds, Parler or Telegram) to target users and generate funds and followers to create their own strategies Maati et al. (2023).
- **State actors:** Governments and institutions can intentionally spread disinformation to influence public opinion or destabilize opponents, both domestically and internationally. Pentney (2022).
- **Political parties:** Organized political entities that utilize disinformation to discredit opponents, shape policy debates, or manipulate public opinion. Hunter (2023).

**Mesosystem: Agents Connecting and Mobilizing Communities**

- **Diasporas:** These communities often play a unique role in transmitting disinformation across borders or between regions, especially when tied to geopolitical or cultural conflicts. In Chernobrov (2022), the author examines the role of diasporas in online conflict narratives, highlighting how the internet has facilitated the creation of "digital diasporas," allowing dispersed communities to connect and negotiate their identities online. Using the 2020 Karabakh war as a case study, Chernobrov (2022) notes how young members of the Armenian diaspora became "cyberwarriors," actively engaging in information dissemination and competing against opposing narratives to shape international perceptions of the conflict.
- **Transboundary Networks:** As Ndahinda and Mugabe (2024) show in the context of ethnic disputes in the Democratic Republic of Congo, digital media—including social platforms—facilitates the emergence of transboundary radical networks. These networks connect militia leaders, supporters, public authorities, customary institutions, civil society actors, and diaspora communities across borders. Through ongoing interactions, these actors collaboratively refine and circulate conspiratorial and hate-filled narratives, often reinforcing violence and deepening social division. These networks illustrate how hybrid communication ecosystems—blending digital and offline spaces—can serve as potent vectors for the amplification and entrenchment of information disorder narratives.
- **Local communities and grassroots movements:** Local-level actors can spread or counter disinformation through interpersonal networks. These actors can reinforce dominant cultural narratives or, conversely, challenge misleading information within their communities. As an example, Kermani (2023) identifies "cyber armies" that operate through eight tactics: "downgrading discussions to the level of the government, justifying the state's policies, cheering up other users, portraying that everything is normal, redirecting debates, spreading fake news, trending misleading hashtags, and mocking dissidents and activists". Nguyen et al. (2023) analyze the Vietnamese American community to better understand how misinformation circulated during the 2016 and 2020 U.S. elections and the COVID-19 pandemic. They focus on grassroots initiatives like VietFactCheck, PIVOT, and The Interpreter, which emerged to counter false narratives. The authors highlight the importance of strengthening local and ethnic media ecosystems to combat misinformation collectively—rather than placing the burden solely on individuals to verify facts.

**Microsystem: Agents Exerting Direct Influence**

- **Influencers:** Individuals who have a large following on social media and can spread disinformation, intentionally or unintentionally, to their followers. Koc Michalska (2023)
- **Political figures:** Politicians or political activists who interact directly with the public, influencing individual beliefs through speeches, social media posts, or personal interactions.

Those figures are framed by Fonseca & Santos Neto (2023) that nowadays politicians become public figures of society who are not public agents anymore, but celebrities and show business personalities that influence the public directly through their social media.

- **Fake news creators (individual entrepreneurs)**: Individuals who generate disinformation for personal or financial gain by creating content designed to manipulate or mislead others. One relevant example is the sock puppets and phone farms agents, Inga et al. (2024) frames it on how this agent's strategy is used among different countries, such as India with the "IT Yoddhas" that works for the Bharatiya Janata Party (BJP) for massive messaging spreading. The authors also cite how Mexico interviewed agents revealed how using "phone farms" runned by one person can run undetected by social media platforms and are able to spread a massive amount of messages on social media without being detected as "a bot".

**2.2 *Raisons d'être***

Disinformation is rarely random. It is a strategic instrument employed by diverse actors to pursue concrete political, ideological, economic, and even psychological goals. Whether the aim is to secure electoral victories, destabilize opponents, deepen societal fractures, or profit from confusion, disinformation functions as a tool for asserting power, influence, and control. We refer to these underlying motivations as *raisons d'être*—a term borrowed from French that literally means "reason for being." We adopt it here because it aptly captures the foundational purpose or driving force behind disinformation strategies. In Table 4, we present an overview of the key *raisons d'être*, accompanied by descriptions, illustrative examples, and the types of agents involved.

These motivations are not static or mutually exclusive. Rather, they emerge from a complex interplay of individual intent, structural opportunity, and strategic alignment. In some cases, unexpected alliances may form among actors who, despite ideological differences, converge around shared objectives. The "Deep State" conspiracy theory, for instance, as analyzed by Berg (2023), illustrates how a narrative originally rooted in a specific political context can be co-opted by a range of actors—political operatives, grassroots movements, conspiracy theorists—each mobilizing it to serve distinct agendas. This flexibility and ambiguity make disinformation a uniquely adaptive mechanism within the broader ecosystem of information disorder.

Understanding these *raisons d'être* is essential not only to identify *who* spreads disinformation, but to grasp *why* they do so—and how these strategic objectives shape public discourse, polarize societies, and undermine democratic processes.

| *Raison d'être* | Description | Specific objectives (examples) | Agents |
|---|---|---|---|
| **Electoral Gain and/or Political Power** | Disinformation is frequently used by political figures and parties to secure electoral victories, discredit opponents, and consolidate political power. This is especially prevalent in competitive or polarized political environments. | - Disinformation spread during elections to weaken political rivals, as seen in the 2016 U.S. presidential election and 2012 South Korean election (Gunther et al., 2019; Keller et al., 2020).<br>- False narratives about election fraud in the U.S. used to undermine electoral trust and destabilize the political landscape (Arceneaux & Truex, 2023), and to question the legitimacy of democratic processes, benefiting specific political elites (Berlinski et al., 2023).<br>- Alt-science proponents and far-right leaders promoted hydroxychloroquine to undermine established health authorities, driven by a mix of personal beliefs and political gains (Casarões & Magalhães, 2021). | Political Figures, Parties, Influencers, Media Outlets |
| **Destabilization of Political Opponents** | Disinformation is often used to disrupt an opponent's campaign, tarnish their reputation, or create divisions within their base. It is deployed to weaken political opponents by undermining their credibility, spreading false accusations, and creating confusion or distrust among their supporters. | - Trump's Disinformation Campaigns: Trump's disinformation strategies during and after his presidency aimed at discrediting political opponents, such as claims of election rigging or corruption by Democrats (Arceneaux & Truex, 2023).<br>- Hillary Clinton's 2016 Campaign: Disinformation efforts targeting Hillary Clinton included fabricated scandals like "Pizzagate" and health-related conspiracy theories, designed to erode trust and support among voters (Gunther et al., 2019; Stabile et al., 2019). | Political Figures and Parties, Influencers, Media Outlets |
| **Sowing of Societal Divides** | Disinformation is often used to deepen societal divides by exploiting existing tensions related to race, religion, class, or political affiliation. The goal is to create or exacerbate "us vs. them" dynamics, polarizing societies and weakening social cohesion. This strategy is particularly effective in destabilizing communities, fostering mistrust, and weakening the collective ability to address shared challenges. | - Populist Movements: Figures like Donald Trump and other populist leaders often employ disinformation to create an "us vs. them" narrative Balfour (2020) pitting their supporters against perceived elites, immigrants, or other marginalized groups.<br>- Russian Disinformation: During the 2016 U.S. election, Russian state-backed operatives used social media to fuel racial tensions, particularly around the Black Lives Matter movement, aiming to deepen societal divides and destabilize American society (Dawson & Innes, 2019).<br>Ethnic Divides in Afghanistan: In post-2001 Afghanistan, elites manipulated ethnic identities through disinformation to enhance their political power, leading to mass actions like bloc voting and deepening ethnic divisions (Sahar & Sahar, 2021). | State actors, Political figures (particularly populist ones), Influencers, Media Outlets |
| **Ideological Promotion and Cultural Hegemony** | Disinformation is used to advance ideological goals, such as social norms, religious, cultural etc. This can involve promoting certain worldviews, discrediting opposition ideologies, or | - Far-Right Groups spreading conspiracy theories to justify violence and create narratives of national threat (Marcks & Pawelz, 2022).<br>- Religious leaders using disinformation to maintain ideological control and influence public moral discourse (Yendell & Herbert, 2022).<br>- United States elites promoting manipulation of identity markers to influence in mass | Far-right groups, nationalist influencers, State Elites, Religious Leaders and Institutions, Media Outlets |

| | reinforcing cultural dominance. | actions (Sahar & Sahar, 2021) | |
|---|---|---|---|
| **Geopolitical Influence and Destabilization** | States, political actors and civil society often engage in disinformation to exert geopolitical influence, destabilize rival states, or consolidate control over domestic populations. These actions can shape global narratives, disrupt alliances, and weaken adversaries. | - The use of disinformation to political destabilization by the United States, focused on undermining democracies (Landon-Murray et al., 2019)<br>- Russia uses disinformation to influence elections and disrupt Western political systems (Bradshaw et al., 2020), erode trust in Western democratic processes (Justwan & Williamson, 2022), between states and the relation state-society (McKay & Tenove, 2021) and influencing political outcomes in Africa (Paterson & Hanley, 2020).<br>- China shapes global narratives to counter criticisms of its policies, particularly concerning human rights and territorial disputes (Bjola & Papadakis, 2020).<br>- Russia and China employ astroturfing and fabricated narratives to manipulate public perceptions and influence international political discourse, such as the Skripal poisoning and Syrian conflict (Zerback & Töpfl, 2022).<br>- Armenian diaspora using disinformation to sway international opinion during conflicts (Chernobrov, 2022). | State Actors (Governments, intelligence agencies), Diasporas, Political Figures, Media Outlets |
| **Economic and Personal Gain** | Some disinformation campaigns are driven purely by the desire for financial profit or personal gain. This can include generating ad revenue, increasing social media following, or direct financial incentives. | - Macedonian Fake News: Youth-driven disinformation industry in Veles focused on profit generation during the 2016 U.S. election (Hughes & Waismel-Manor, 2021).<br>- Social Media Influencers: Spreading disinformation for visibility and monetization during the COVID-19 pandemic (Torreblanca, 2023).<br>- Social media platforms benefiting from increased engagement through the spread of sensationalist and false content (Albertson, 2020). | Individual 'Entrepreneurs' (Fake news creators, social media influencers), Social Media Platforms (benefiting from increased user engagement), Media Outlets |

*Table 4: Systematized Description on "Raison d'être" of Agents of Information Disorder. Source: elaborated by the authors.*

## 3. Tactics: Crafting Messages, Exploiting Platforms, Manipulating Flows, and Infiltrating Media

This third stage of the information disorder cycle examines the diverse tactics employed by agents to achieve their objectives. These tactics exploit sociopolitical and informational vulnerabilities specific to a given context, which were identified through the coding of articles under the categories "Strategies of Legitimation" (number of segments codified: 81), "Specific Topics of Disinformation Addressed" (number of segments codified: 187), "Tactics for Spreading Disinformation" (number of segments codified: 459) and "Type of Media or App Studied" (number of segments codified: 177). While prior literature has explored the factors that facilitate the spread of information disorder, this section builds on that work by emphasizing strategies of legitimation—those aimed at enhancing the credibility and persuasive power of false or misleading messages in the political or electoral arena. In essence, these tactics are designed to maximize both the reach and impact of disinformation. While many rely on the affordances of the Internet, social media, and other technologies, their effectiveness extends beyond technological means.

This analysis is informed by Foulger's Ecological Model of Communication, which identifies four key components of any communicative act: the message, the language used to construct it, the medium through which it is transmitted, and the people who create and consume it (Chew & Ng, 2021). Drawing from these components and patterns observed in our empirical analysis, we classify tactics into four broad categories, that span from macro-level strategies affecting the information ecosystem to micro-level tactics that shape user engagement. **At the micro level, these include (1) crafting persuasive and credible-seeming messages, and (2) exploiting platform affordances. At the macro level, they encompass (3) manipulating the flow of information, and (4) infiltrating the media ecosystem**. Each category and its corresponding tactics is explored in the following sections.

A critical characteristic of these strategies is their iterative nature. As Hughes and Waismel-Manor (2021) note, the expertise demonstrated by information disorder agents often emerges through a "process of trial and error." Tactics are continuously refined and updated, frequently incorporating new technologies such as deepfakes to enhance impact (Paterson & Hanley, 2020). Not all campaigns deploy the full array of tactics, and even when false or misleading information is shared unintentionally, the results can significantly disrupt communication processes, amplify harmful narratives and negatively impact democratic governance and institutions.

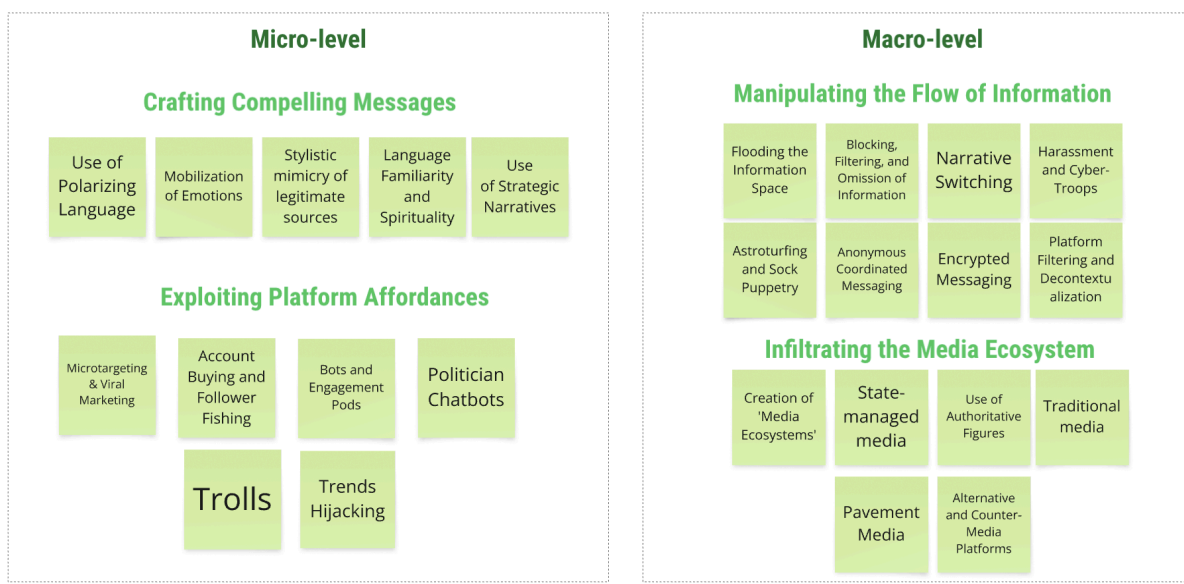

*Figure 5: Systematized Tactics. Source: elaborated by the authors.*

Importantly, this framework does not view audiences as passive recipients. Instead, individuals actively interpret and negotiate the meaning of messages based on their experiences and sociopolitical contexts. This dynamic underscores ongoing scholarly debates about the behavioral and attitudinal impacts of misinformation—topics addressed more fully in the next stage of the framework.

**3.1 Crafting Compelling Messages**

Tactics for crafting messages aim to ensure they are not only believable but also emotionally resonant and engaging. These narratives must appear legitimate while also appealing to audiences in ways that provoke strong reactions or align with preexisting beliefs. Crafting such messages involves both content and form—language choices, writing style, formatting, and the use of culturally or politically salient cues.

Our analysis identifies four key tactics under this category:

- **Use of Polarizing Language:** Language plays a central role in shaping public opinion (Albertus & Makoza, 2023). Its "inherent malleability" allows political actors to weaponize words, transforming opinions into apparent facts and shaping perceptions of reality (Baron & Ish-Shalom, 2024). This tactic exploits language's capacity to shift meaning, particularly by leveraging divisive issues to inflame societal tensions. A prominent example is the Russian-backed Internet Research Agency, which employed polarizing rhetoric to deepen political divides in the United States, often by vilifying opposing political or social groups (Nonnecke et al., 2022). Similarly, in the eastern Democratic Republic of Congo, anti-Banyamulenge and anti-Tutsi hate speech has incited real-world violence (Ndahinda & Mugabe, 2024). Polarization is often intensified through the stylistic elements of alternative media. Bjola and Papadakis (2020) note that such outlets frequently rely on sensationalist headlines and highly partisan content to spread falsehoods under the guise of journalism. Furthermore, the rhetorical strategies used in these narratives frequently draw on inflammatory or emotionally charged language—an aspect further examined in the following section—that deliberately reinforces social cleavages and entrenches polarized identities (Ndahinda & Mugabe, 2024; Campos & Barili, 2022). Russian state-sponsored media outlets such as RT and Sputnik have exemplified this tactic by persistently portraying Black Lives Matter (BLM) protesters in a negative way, thereby stoking racial tensions in the United States (Bradshaw et al., 2023). These examples underscore how language serves as a powerful tool in crafting information disorder narratives that are both emotionally resonant and ideologically divisive.
- **Mobilization of Emotions**: Information disorder narratives frequently exploit emotional responses—particularly anger, fear, hatred, and anxiety—to increase message receptivity and drive engagement (Alberton & Guiler, 2020; Edwards III, 2020; Nonnecke et al., 2022; Marcks & Pawelz, 2022). Anger, for instance, can provoke moral outrage and a sense of urgency, prompting individuals to share or engage with content more rapidly. Fear amplifies perceived threats, making individuals more susceptible to manipulation, while hatred is often used to further entrench in-group/out-group dynamics. Although negative emotions dominate most information disorder strategies, positive emotional appeals are also strategically deployed. Media outlets such as Russia-linked RT have employed humor and emotionally resonant storytelling to build affinity with target audiences, thereby enhancing their legitimacy at both domestic and international levels (Bradshaw et al., 2023). As Bradshaw et al. (2023) argue, such emotional appeals function as a form of soft-power persuasion—aimed at co-opting rather than coercing audiences by tapping into shared anxieties, values, or aspirations. Ultimately, emotional mobilization operates not only as a tool for engagement but also as a mechanism for reinforcing social fragmentation and facilitating the spread of disinformation and misinformation.
- **Stylistic mimicry of legitimate sources:** A widely used tactic in the spread of information disorder involves imitating the stylistic elements of credible news sources to lend legitimacy to false or misleading content. Fake news websites often replicate the visual identity of established media outlets—adopting familiar fonts, layouts, and design conventions—to foster an appearance of authenticity. These sites may include genuine news items, such as weather updates or sports coverage, interspersed with disinformation that is frequently xenophobic, conspiratorial, or otherwise manipulative (Torreblanca, 2023). According to Torreblanca (2023), this tactic enables such websites to operate as "repositories and launderers of fraudulent information." Although they mimic the structure and tone of journalistic organizations, these sites often serve the interests of specific political actors or causes, functioning primarily to disseminate false narratives under the guise of legitimate reporting. There are, however, multiple forms of stylistic mimicry beyond website aesthetics. These range from the use of "visual evidence collages" on fringe forums like 4chan to construct visual narratives (Krafft & Donovan, 2020), to the dissemination of false health advisories during the 2020 Israeli elections via WhatsApp, which employed official logos and formatting to create a veneer of institutional credibility (Haleva-Amir, 2022). These examples underscore the persuasive power of stylistic cues in reinforcing trust in deceptive narratives.
- **Language Familiarity and Spirituality:** Information disorder narratives also gain legitimacy through cultural familiarity and spiritual resonance—especially among diaspora communities. Nguyen et al. (2023) document how older Vietnamese Americans frequently turn to Vietnamese-language YouTube channels that translate and rebroadcast right-wing narratives from outlets like Breitbart, OAN, Fox News, or Epoch Times. These creators, incentivized by monetization structures tied to views and subscriptions, embed political disinformation in formats that feel accessible and trustworthy to their audience. In some cases, content is framed using religious language or emotionally evocative phrases, such as "thank God, Trump is gonna destroy the China Communist regime," which references both anti-communist sentiment and spiritual conviction. These tactics blend linguistic familiarity, shared histories, and religious narratives to enhance the persuasive power of misleading messages, making them more likely to be accepted and internalized.
- **Strategic Narratives:** Strategic narratives are context-specific stories designed to exploit sociopolitical vulnerabilities. These narratives embed misinformation, propaganda, conspiracy theories, or hate speech within compelling rationales. For instance, falsehoods linked to the COVID-19 pandemic, racial unrest, or armed conflict often frame one group as unjustly suffering while another benefits. This antagonistic logic enhances resonance and reinforces in-group/out-group dynamics. These narrative strategies include:
  - **Election fraud claims:** "A powerful group of people is acting illegally and lying to 'you'. They are threatening democracy. Don't let them."
  - **Large-scale conspiracy theories** (e.g., Deep State, QAnon): "Everything is manipulated behind the scenes by hidden elites. Only the initiated can see the truth."
  - Discrediting political opponents: "That candidate is a threat to your country, your morals, your family. They are dangerous."
  - **Scapegoating and racialized blame**: "'They' are out to get 'you'. They are responsible for your suffering, and must be stopped."
  - **Hate speech and incitement to violence**: "'They' are subhuman, corrupt, or immoral. 'We must defend ourselves."
  - **Anti-establishment narratives**: "You can't trust the media, politicians, or scientists—they're all in on it. Only we speak the truth."
  - **Anti-science narratives**: "Science is corrupt and politicized. Experts don't care about 'your' health or freedom—take back control."
  - **Geopolitical manipulation**: "Foreign powers are undermining our sovereignty, values, and way of life. We must resist their influence."
  - **Ideological promotion and cultural hegemony**: "Our worldview is under attack. Only by embracing 'our' values can society be saved."
  - **National threat framing**: "The nation is under siege—by outsiders, traitors, or enemies

within. Extraordinary measures are justified.

## 3.2 Exploiting Platform Affordances

Digital media platforms—ranging from open networks like Facebook, YouTube, and X (formerly Twitter) to encrypted apps such as WhatsApp and Telegram—have become central channels for the amplification of information disorder (Bradshaw et al., 2023; Inga et al., 2024). These platforms enable increasingly sophisticated strategies that exploit user behavior, algorithmic incentives, and interface design to enhance the visibility and perceived legitimacy of disinformation. This section examines tactics that shape how users engage with disinformation. While they often serve broader political or ideological goals, these tactics are designed to boost reach, traction, and influence through interaction—whether by gaming platform metrics, provoking engagement, or creating the illusion of popularity.

- **Microtargeting and Viral Marketing:** Digital platforms facilitate the microtargeted delivery of disinformation to specific audience segments, based on behavioral and demographic data. This selective targeting allows agents of information disorder to influence and mobilize groups with tailored narratives (Vićić & Gartzke, 2024). Viral marketing builds on this by designing content—such as images, videos, or deepfakes—that is emotionally provocative and easily shareable. Recommendation algorithms further reinforce exposure by prioritizing this type of content (Campos & Barili, 2022). Russian operatives, for example, reportedly reached over 120 million Americans via Facebook during the 2016 U.S. election (Haleva-Amir, 2022).
- **Account Buying and Follower Fishing**: Follower fishing involves rapidly following large numbers of users to gain reciprocal follows, thereby increasing the apparent credibility of the original account. Once follow-backs are secured, the account unfollows others to boost its followers-to-following ratio (Dawson & Innes, 2019). In parallel, account buying bypasses the slow process of audience-building by purchasing pre-established accounts (Dawson & Innes, 2019). These can be used to inject disinformation into existing follower networks, often marketed by websites offering thousands of followers for a nominal fee. Designed "to build audience and influence", this technique has been employed by state-linked groups such as Russia's Internet Research Agency to promote Russia's agenda against Germany and the UK during Brexit (Dawson & Innes, 2019).
- **Bots and Engagement Pods:** Bots—automated or semi-automated social media accounts—are widely used to inflate engagement metrics and simulate popularity. Their functions range from simple tasks (e.g., liking or retweeting) to sophisticated AI-driven content generation (Nonnecke et al., 2022). Political bots have been central to computational or digital propaganda, notably in the 2016 U.S. presidential election and the 2018 U.S. midterm elections, where they were used to fabricate consensus on issues such as immigration (Haleva-Amir, 2022; Nonnecke et al., 2022). Recent developments include "engagement pods," hybrid networks of bots and human users that artificially boost visibility through coordinated interactions (Haleva-Amir, 2022).
- **Politician Chatbots:** A more advanced use of bot automation was observed in the 2019 Israeli elections, where political leaders—including Prime Minister Netanyahu—deployed Facebook Messenger chatbots. These bots microtargeted voters based on their responses during interactive sessions, collecting data on political preferences without users' consent and conducting hidden canvassing (Haleva-Amir, 2022). The "Bibi-Bot" engaged users with games and entertainment, while simultaneously gathering voting data (illegally) to refine campaign messaging. Parallel investigations uncovered a network of hundreds of pro-Netanyahu fake accounts that disseminated over 130,000 tweets in the months leading up to the election (Haleva-Amir, 2022).
- **Trolls**: In the context of information disorder, trolls operate as attention hijackers, injecting divisive or distracting content into comment threads, hashtags, or news cycles to drive engagement and visibility. By prompting outrage or confusion, they exploit the attention economy of digital platforms, pushing manipulated narratives into algorithmic relevance. Their activity reinforces polarization, increases content virality, and creates the illusion of broad public support or controversy around specific topics (Ortiz, 2020, as cited in Nonnecke et al., 2022). In this way, trolling is not just disruptive—it is performative amplification designed to manipulate user interaction patterns.

- **Trends Hijacking**: Information disorder agents often exploit trending algorithms to elevate disinformation into mainstream visibility. Platforms like X (formerly Twitter) enable this through trending topics and hashtags, which can be manipulated via coordinated bot and troll activity (Rasheed & Naseer, 2021). This allows fringe or false narratives to enter the public sphere and shape debate. According to Rofrío et al. (2019), this tactic was used during the 2017 presidential elections of Ecuador to create false trending topics.
- **Encrypted Messaging**: Encrypted messaging platforms such as WhatsApp, Telegram, and Signal have become potent vectors for disinformation, offering high user privacy and limited traceability. End-to-end encryption ensures only intended recipients can access content—shielding falsehoods and hate speech from public scrutiny and regulatory oversight, as seen during Brazil's 2018 presidential elections (Lôbo & Bolzan de Morais, 2019). This opacity benefits state-aligned propagandists and authoritarian actors aiming to shape discourse beyond the reach of traditional media and fact-checkers (Inga et al., 2024). These platforms thus contribute to a fragmented, unmoderated information ecosystem where disinformation circulates unchecked.

## 3.2 Manipulating Flows of Information

While the previous section focused on tactics that exploit platform affordances to amplify disinformation, this category centers on strategies that distort the flow, structure, and accessibility of information itself. Rather than spreading content for visibility alone, these tactics aim to destabilize the broader informational ecosystem and to create a perception of "unruly communication environment", as described by Jungherr and Rauchfleisch (2024), while also eroding trust in institutions, particularly the media, and advancing political agendas. Scholars have referred to this phenomenon as contributing to a post-truth society or state of "epistemic anarchy" (Dawson & Innes, 2019), where citizens struggle to discern credible information amid competing and often contradictory narratives.

These strategies often rely on weak regulatory enforcement and lack of transparency mechanisms (Haleva-Amir, 2022). Notably, they do not always involve outright falsehoods. Instead, they operate through more subtle techniques—limiting access, fragmenting contexts, overloading with distraction, or artificially stimulating consensus.

- **Flooding the Information Space:** This tactic involves saturating the public sphere with excessive, irrelevant, or misleading content to obscure critical narratives and divert attention from contentious issues (Cirone & Hobbs, 2023). Commonly employed by authoritarian regimes and political actors across the ideological spectrum, flooding aims to dilute the visibility of dissent, destabilize the flow of verified information, and exhaust public attention. Social media bots and trolls are frequently the tools of choice used to overwhelm platforms with a mix of neutral, banal, or distracting content—blending into discourse while simultaneously amplifying divisive messages, drowning out opposition, and distorting public debate (Grover et al., 2019; Marlow et al., 2021). Empirical examples illustrate how this tactic is weaponized in practice. In China, government-affiliated users have been shown to disseminate fabricated posts en masse to deflect online discussions away from sensitive topics (King et al., 2017, as cited in Cirone & Hobbs, 2023). Similarly, during Venezuela's 2014 anti-Maduro protests, authorities used Twitter to flood timelines with apolitical messages, effectively muting criticism and impeding mobilization by the opposition (Cirone & Hobbs, 2023).
- **Blocking, Filtering, and Omission of Information**: The inverse of flooding, this approach entails the systematic exclusion or suppression of relevant information. Authoritarian governments like those in China and Russia often block access to foreign news sources to control public opinion and suppress dissent (Torreblanca, 2023). This control strategy is not exclusive to

- autocracies. In Mexico, local governments have limited citizen access to information during policy deliberations on large infrastructure projects, simulating transparency while eroding public trust (Díaz Aldret, 2018).
- **Narrative Switching**: Narrative switching refers to the gradual or abrupt redirection of messaging by disinformation agents. Initially, campaigns appear neutral or mundane, aligned with a persona's expected tone. Over time, these accounts pivot toward politically charged content—often aligned with shifting geopolitical goals. For example, the Internet Research Agency (IRA) moved from banal content to pro-Russian and polarizing messages depending on operational priorities (Dawson & Innes, 2019). The effect is not just amplification, but confusion and destabilization of the informational ecosystem.
- **Harassment and Cyber-Troops:** Harassment campaigns—often coordinated and sustained—aim to intimidate and silence critics, journalists, researchers, and public figures who challenge disinformation. These attacks, frequently launched by troll networks, degrade trust in digital platforms as spaces for open dialogue and dissent. Sometimes orchestrated by state-affiliated "cyber troops" or informal networks of supporters, these campaigns contribute to a hostile discursive environment and undermine civic participation. Documented examples include Mexico, Azerbaijan, and South Korea, where targeted harassment has had measurable chilling effects on political debate and media freedom (Nonnecke et al., 2022). The result is a fragmented and toxic information ecosystem where falsehoods spread with reduced resistance. The blending of human and automated actors further obscures accountability and scale. On Reddit, for example, just 1% of users were found responsible for 74% of conflict-related interactions—underscoring the disproportionate influence of coordinated harassment on digital discourse (Haleva-Amir, 2022).
- **Astroturfing and Sock Puppetry:** These tactics simulate grassroots support through centrally coordinated campaigns involving fake personas or communities. By manufacturing consensus and silencing dissent, they distort public perception of legitimacy or popularity (Zerback & Töpf, 2022; Haleva-Amir, 2022; Keller et al., 2019). The 2012 South Korean presidential election offers a clear example, where the National Intelligence Service orchestrated an astroturfing campaign to influence voter behavior (Keller et al., 2019). Distinguishing genuine activism from manipulation becomes difficult, particularly when real users are co-opted into coordinated efforts.
- **Anonymous Coordinated Messaging:** Anonymous disinformation campaigns also play a significant role in eroding trust during key democratic processes. During Israel's 2019 and 2020 elections, spammed voice and text messages—circulated without attribution—misled voters, defamed candidates, and spread alarmist appeals (Haleva-Amir, 2022). These tactics persisted despite legal restrictions, revealing enforcement gaps in regulating electoral disinformation.
- **Platform Filtering and Decontextualization:** Krafft and Donovan (2020) describe how digital platforms enable content to be removed from its original conversational context, such as threads or comment chains. As content circulates across fragmented spaces, critical contextual cues—such as tone, authorship, or prior rebuttals—are lost. While enabled by digital media features, this process is not aimed to amplify disinformation, but to facilitate its persistence and misinterpretation, reinforcing fragmented and polarized information environments.

### 4.3. Infiltrating the Media Ecosystem

The final category of tactics targets the long-term shaping of public discourse by embedding information disorder into the very fabric of media and communication systems. Unlike amplification or disruption strategies, which focus on reach or destabilization, these tactics aim to normalize and legitimize specific narratives over time—ultimately influencing political attitudes, voting behavior, and broader patterns of civic engagement. Their effectiveness hinges not only on the content of the message but also on the credibility of the platforms, voices, and institutions through which they are delivered.

These strategies are employed by a wide range of actors, including state-aligned media, political campaigns, alternative platforms, and even traditional news outlets. As Inga et al. (2024) note, this gradual infiltration of discursive spaces contributes to what they term *informational autocracy*—a condition in which governments (and, as we argue, other actors) maintain legitimacy while manipulating public opinion through tightly controlled yet outwardly pluralistic media environments.

- **Creation of 'Media Ecosystems'**: Information disorder agents often build closed media ecosystems that appear credible and community-driven while systematically excluding dissent. These ecosystems—frequently established through platforms like WhatsApp or Telegram—are rooted in affinity, trust, and repetition. Dissenting voices are filtered out, creating ideological "fiefdoms" where only reinforcing narratives circulate (Solano Gallego, 2018, as cited in Lôbo & Bolzan de Morais, 2019). Brazil's 2018 presidential election illustrated this approach, with coordinated WhatsApp campaigns leveraging fear and superstition to undermine traditional sources of authority, including science and journalism.
- **State-managed media:** Authoritarian and democratic governments alike have used state-owned or state-affiliated media to subtly embed disinformation into public discourse. Outlets like Russia's RT or China's CGTN cultivate legitimacy as professional international broadcasters while using that credibility to sow distrust in democratic institutions or amplify polarizing narratives (Bradshaw et al., 2023). Their strength lies not only in reach, but in blurring the line between legitimate news and strategic influence.
- **Use of Authoritative Figures**: Political figures with high visibility and institutional power can serve as vectors of disinformation and misinformation, lending credibility through their perceived authority. For example, U.S. President Donald Trump's minimization of COVID-19 risk following his recovery served not only to spread misinformation but to reframe public discourse around the pandemic (Albertus & Makoza, 2023). These figures have the power to position themselves as "arbiters of truth" (Balfour, 2020), undermining journalistic norms and polarizing the informational landscape.
- **Traditional media:** Despite growing distrust in mainstream outlets, traditional media—TV, radio, print—continue to play a pivotal role in shaping public understanding. Jungherr & Rauchfleisch (2024) further highlight that "disinformation reaches people predominantly through media coverage and public discourse rather than digital media." In countries like Ghana, traditional platforms have been used to legitimize political disinformation by hosting partisan commentators, known locally as *Sojojin Baci* or "soldiers of the mouth" (Gadjanova et al., 2022). Often, disinformation circulates between digital and traditional spaces: social media rumors gain traction offline through radio talk shows or newspaper commentary, reinforcing their reach and credibility (Gadjanova et al., 2022).
- **Alternative and Counter-Media Platforms:** So-called alternative or "counter-media" outlets are independent from state control but often challenge the legitimacy of mainstream journalism by promoting disinformation, conspiracy theories, or partisan narratives. While some, like Peru's *Canal N*, have played roles in holding power accountable, others—such as Breitbart in the U.S., MV-Lehti in Finland, or Russia' The NOW and Redfish—foster epistemic distrust and polarization (Li et al., 2022; Bjola & Papadakis, 2020; Bradshaw et al., 2023). These platforms often mix factual content with misleading or false claims and are frequented by audiences disillusioned with traditional media. Many of these platforms are run by disenchanted amateurs (Hiltunen, 2018, as cited in Bjola & Papadakis, 2020). Nonetheless, they serve as fertile ground for the indirect circulation of state-backed disinformation, especially around polarizing "trigger topics" such as immigration, gender, or religion (Bjola & Papadakis, 2020).
- **Pavement Media:** Pavement media refers to informal communication about public affairs that takes place in everyday offline settings such as markets, places of worship, bars, and community gatherings (Gadjanova et al., 2022). Though non-digital in form, these spaces serve as powerful amplifiers of disinformation by circulating narratives through trusted interpersonal

networks—particularly among indirect social media users. Their perceived legitimacy often heightens their influence over public opinion, policy debates, and democratic processes. In Nigeria, for instance, a rumor originating on WhatsApp—that President Buhari had been replaced by a clone—gained traction when it was repeated in radio shows and sermons, including by a prominent preacher addressing a large congregation (Cheeseman et al., 2020, as cited in Gadjanova et al., 2022). This example illustrates how digital-origin disinformation can be absorbed into culturally resonant offline discourse. By blurring the boundaries between online and offline spheres, pavement media embeds information disorder narratives in the everyday flow of public conversation—complicating both detection and mitigation efforts.

**4. Impact and Transformations on the Information Environment**

The information ecosystem has undergone profound transformations over the past decade, not all of which can be directly attributed to disinformation alone. That being said, while social media has played a role in democratizing information and facilitating civil revolutions, it has also created new vulnerabilities in the information environment, making them susceptible to assaults through the internet (Vićić & Gartzke, 2024).

Our analysis identified multiple ways in which information disorder is undermining the information environment. These findings emerged from systematic coding under the category "Impact of Disinformation" (371 segments). We adopt the term "information disorder" instead of narrowly referring to "disinformation" or "misinformation," aligning with Ireton & Posetti (2018), because it captures the broader systemic disruptions caused by deceptive information practices—not only intentional falsehoods (disinformation) but also unintended inaccuracies (misinformation), contextual distortions, and the structural factors enabling their spread.

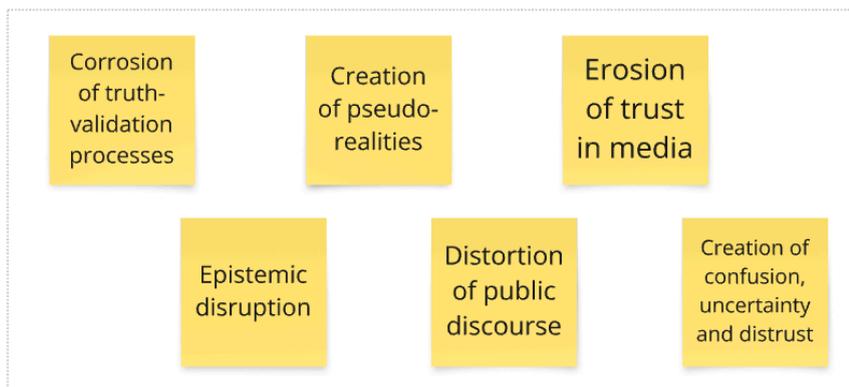

*Figure 5: Impacts of Disinformation. Source: elaborated by the authors.*

The distortion of public discourse, erosion of trust in credible sources, damage to the public sphere and rational deliberation, and creation of an atmosphere of confusion and uncertainty are largely agreed upon impacts. Specific impacts highlighted in our analysis include:

- **Corrosion of truth-validation processes:** Disinformation actively undermines the ability of societies to collectively validate truth claims, severely deforming the public sphere and fostering radicalization and political polarization (Bjola & Papadakis, 2020). This disruption of societal consensus around basic facts facilitates the proliferation of conspiratorial and extremist narratives.
- **Epistemic disruption:** In a state of information disorder, the distinctions between facts and subjective opinions are blurred, creating confusion and severely reducing a society's capacity for informed democratic decision-making (Baron, 2024). In such an environment, fact-based debates become challenging, making collective solutions elusive.

- **Creation of pseudo-realities:** Disinformation creates environments dominated by opinion rather than fact, disabling critical judgment and fostering closed, self-reinforcing opinion bubbles (Baron, 2024).
- **Distortion of public discourse**: Disinformation shifts public debate towards sensational, polarizing, or emotionally charged topics. This shift is amplified by social media algorithms designed to prioritize engagement over informational accuracy, the basis of the "attention economy." As a result, sensationalist or confrontational content overshadows nuanced, evidence-based discussions on critical policy issues (Baron, 2024; Invernizzi & Mohamed, 2023).
- **Erosion of trust in media**: One of the most significant impacts is the erosion of public trust in the media and the information environment broadly. Studies have shown a dramatic decline in media trust over the past two decades. According to Torreblanca (2023), citing Gallup Institute data, trust in media among Americans declined from 54% in 2003 to 32% by 2016, primarily due to perceived inaccuracies and biases. Li et al. (2022) further explain that alternative and fringe media sources frequently delegitimize mainstream media by framing them as politically manipulated or inherently biased.
- **Creation of confusion, uncertainty and distrust**: Persistent exposure to information disorder leads to confusion (citizens are unsure about what is true and what is not), fostering widespread skepticism and uncertainty (Invernizzi & Mohamed, 2023). Such an environment not only makes it difficult to agree on basic realities, but also systematically erodes public trust not just in media but in democratic institutions, scientific expertise, and governmental authorities as well. This erosion is notably intensified by tactics such as information flooding, where deliberate saturation with misleading content makes it impossible for ordinary citizens to navigate informational accuracy confidently (Bjola & Papadakis, 2020).

**5. Impacts on Society, Policy and Democracy at Large**

This final stage examines the tangible consequences of disinformation on broader societal structures, democratic governance, and public policy. Beyond reshaping the information environment, scholars have documented how information disorder disrupts societies, alters political dynamics, and impacts democratic institutions. Disinformation does not merely influence individual beliefs and behaviors; it profoundly affects political decision-making processes, social cohesion, and democratic stability.

It is worth mentioning that the studies selected for our systematic literature review exhibit substantial methodological diversity, encompassing quantitative, qualitative, experimental, and mixed-method approaches (Code "Methodological Approach", number of segments codified: 380). Quantitative methodologies include panel surveys (Arceneaux, 2023; Freiling et al., 2023), survey experiments employing randomized treatments (Albertson, 2020; Enders, 2019), and automated text analysis coupled with manual coding of social media datasets (Cirone & Hobbs, 2023; Bradshaw et al., 2023). Qualitative methods are also well represented, with studies employing critical discourse analysis to examine media and political texts (Albertus & Makoza, 2022), semi-structured interviews and focus groups to gain deeper insights into individual attitudes and behaviors (Chernobrov, 2022; Gadjanova et al., 2022), and ethnographic research combined with document analysis (Christensen, 2022). Additionally, mixed-method approaches, integrating both survey data and qualitative analyses (Bjola & Papadakis, 2020), demonstrate an increasing effort among scholars to triangulate findings for greater robustness and contextual depth.

Given this methodological diversity, it is important to highlight that our review is not limited to studies employing strictly quantitative or positivist approaches designed to establish direct causality. Indeed, assessing the impact of disinformation is particularly challenging. According to our review, there is no consensus on the tangible impacts of disinformation, and particularly in the context of public policy – which may be due to different scopes and methodologies to target the question. This issue is particularly challenging due to the multifaceted nature of disinformation and public policy, which makes it difficult to understand the constitutive relationship between the two.

To systematize our findings, we employ Klaus Frey's conceptual framework distinguishing among **policy, politics, and polity** (Frey, 2000). According to Frey:

- **Politics** refers to the dynamic processes involving political negotiations, power relations, electoral competition, and partisan debates.
- **Policy** denotes the outputs or outcomes of governmental decision-making—specific decisions, actions, and programs enacted by governments.
- **Polity** encompasses the institutional and normative structures underpinning democratic governance, including democratic norms, checks and balances, rule of law, and the functioning of governmental institutions.

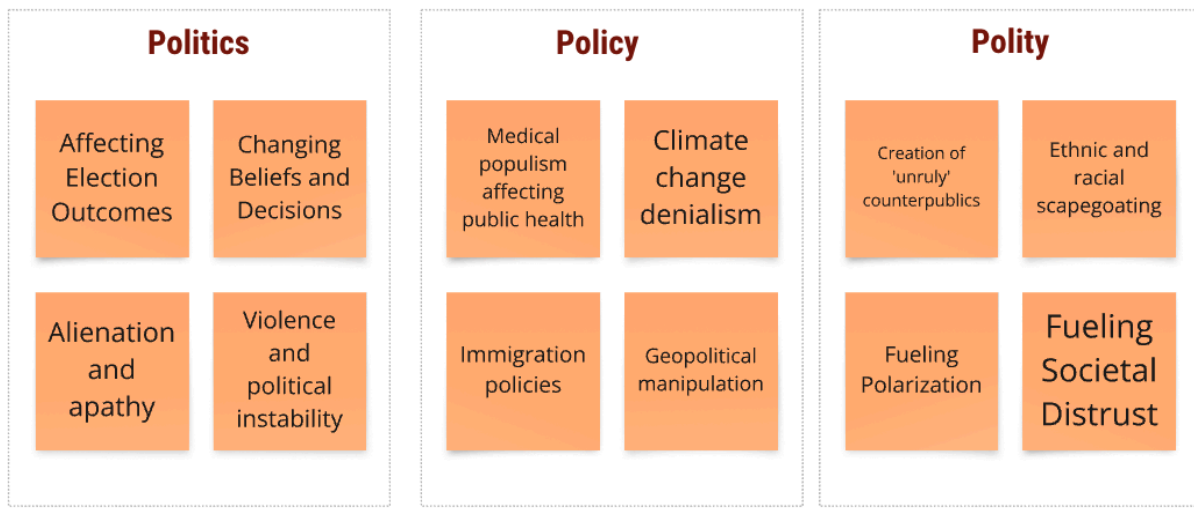

*Figure 6: Systematized findings. Source: elaborated by the authors, based on Klaus Frey's conceptual framework for Politics, Policy and Polity. (Frey, 2000)*

Leveraging this distinction allows us to articulate how disinformation affects democratic processes and democracy at large: it operates directly on the realm of politics, affecting electoral outcomes and political competition. Simultaneously, disinformation indirectly—though intentionally—impacts policy and polity, influencing the substance of governmental actions and eroding foundational democratic norms and institutions. Concretely, our analysis identified multiple societal impacts through systematic coding under the category "Impact of Disinformation" (371 segments). These impacts can be classified using Frey's framework as follows:

**Politics – Impacts on Political Processes and Democratic Participation**

Disinformation directly undermines key democratic processes, particularly electoral processes. For instance, disinformation concerning electoral fraud erodes public trust in election outcomes, fueling political instability and polarization (Albertus & Makoza, 2022; Hunter, 2023). Furthermore, narratives delegitimizing political opponents exacerbate partisan conflict, undermining pluralistic debate and democratic tolerance (Bjola & Papadakis, 2020; Hidalgo, 2022).

- **Affecting Election Outcomes**: Disinformation's influence on elections is one of the most studied areas, with mixed results regarding its actual impact. Paterson & Hanley (2020) argue that disinformation can cause voters to question democratic processes, decrease trust in election results, and exacerbate political tensions, ultimately leading to the erosion of political legitimacy. Mauk & Grömping (2024) found that disinformation contributes to less accurate beliefs about election fairness, amplifies polarization, and undermines trust in electoral integrity, affecting both public perceptions and election outcomes. Although the full extent of disinformation's ability to sway elections remains debated, evidence suggests it can play a significant role in shaping electoral outcomes, particularly in tightly contested or polarized environments. Empirical examples include Facebook's role in spreading conspiracy theories in Zimbabwe (Karekwaivanane, 2019), Telegram's role during the Ukraine conflict (Inga et al., 2024), and WhatsApp's role in disseminating false narratives during the 2018 Brazilian elections (Lôbo & Morais, 2019). Additionally, studies suggest that the Internet Research Agency's campaigns via Twitter and Facebook significantly influenced the U.S. electoral climate (Albertus & Makoza, 2022).

- **Changing Beliefs and Decisions**: The literature shows that disinformation can influence individual beliefs and decisions, though the extent of this influence is debated. For instance, during the 2017 German parliamentary election, exposure to disinformation significantly impacted vote choice, particularly driving voters towards the right-wing populist AfD and away from the mainstream CDU/CSU (Zimmermann, 2020). This suggests that while disinformation may not broadly alter public opinion, it can effectively shift the political allegiances of targeted groups, especially those predisposed to distrust mainstream media and politics. Similarly, disinformation during the Russia-Ukraine war, as studied by Zhao et al. (2024), was found to manipulate public opinion, polarize society, and undermine trust in institutions, exacerbating conflicts and contributing to societal fragmentation.
- **Alienation and apathy**: Exposure to information disorder, particularly to conspiratorial narratives is associated with political alienation and reduced civic participation, thus weakening democratic responsiveness (Czech, 2022).
- **Violence and political instability:** Disinformation campaigns have been widely associated with violent political events like the January 6th insurrection, fueled by falsehoods propagated by influential figures (Albertus & Makoza, 2022).

**Policy – Consequences for Public Policy and Decision-making**

One of the key questions is whether disinformation can extend its influence to impact policy-making. Perl et al. (2018) suggests that disinformation disrupts evidence-based policy-making by promoting alternate realities and selective information, leading to policy decisions based on misleading evidence and undermining established policy processes. Though indirect, disinformation also may impact policy by shaping both public opinion and policymaker perceptions. The findings below highlight the potential for disinformation to not only shape public opinion but also to directly impact the policy decisions that affect society as a whole.

- **Medical populism affecting public health:** Leaders advocating discredited medical treatments, such as hydroxychloroquine during COVID-19, leveraged disinformation to undermine public health recommendations and scientific authority (Casarões, 2024). This complicated public health policymaking and undermined scientific consensus, directly harming public health outcomes (Casarões, 2024).
- **Climate change denialism:** Marlow et al. (2021) found that bots spreading disinformation on climate change exacerbate polarization, promote denialist views, and distort public discourse on environmental issues.
- **Immigration policies:** Disinformation narratives regarding migration and national security similarly affect immigration policies and public spending decisions, often leading to restrictive and exclusionary measures (Heckler & Ronquillo, 2019).
- **Geopolitical manipulation**: Strategic disinformation narratives shape international perceptions, intensify geopolitical rivalries, and destabilize domestic governance (Bjola & Papadakis, 2020). Similarly, Landon-Murray et al. (2019) note that disinformation in U.S. foreign policy erodes political accountability, destabilizes targeted regions, and undermines democracy by influencing public perceptions and policy decisions.

**Polity – Erosion of Democratic Institutions and Normative Foundations**

Finally, disinformation profoundly impacts the broader institutional and normative foundations of democracy, weakening societal consensus on democratic values and institutional legitimacy. Persistent exposure to information disorder not only erodes public trust in media and information (as mentioned above), but also in governmental institutions, and expertise-based institutions (Torreblanca, 2023; Baron, 2024). This institutional distrust is associated with decreasing civic engagement, and increasing political alienation, which in turns fosters environments conducive to authoritarian tendencies (Zimmermann, 2020; Czech, 2022).

- **Creation of 'unruly' counterpublics:** Disinformation facilitates the emergence and mobilization of anti-democratic counterpublics—subgroups within society that position themselves in opposition to mainstream democratic values and public discourse. By systematically reinforcing and amplifying narratives that undermine democratic institutions, these counterpublics foster identity-based polarization and radicalize segments of the population. Bjola and Papadakis (2020) illustrate how strategically disseminated disinformation strengthens extremist communities, legitimizes anti-establishment rhetoric, and promotes divisive identity politics, ultimately challenging social cohesion and democratic stability. For instance, platforms like Telegram and Gab have provided fertile ground for extremist communities (e.g., white nationalists, QAnon supporters) to form, spread conspiratorial narratives, and engage in coordinated political action that opposes democratic norms and institutions.
- **Ethnic and racial scapegoating:** Historically, racist and xenophobic disinformation narratives have been instrumentalized to justify exclusionary and oppressive policies, severely impacting marginalized communities. Heckler and Ronquillo (2019) demonstrate how misinformation depicting racial and ethnic minorities as threats or scapegoats for broader societal problems has historically justified harmful political agendas, restrictive immigration policies, and even acts of violence. For instance, anti-immigrant disinformation campaigns in the U.S. and Europe have portrayed migrant populations as criminals or economic burdens, directly influencing immigration legislation and public attitudes toward migrants and refugees. Similarly, targeted disinformation against ethnic groups, such as the Rohingya in Myanmar, has been explicitly linked to severe human rights abuses, forced displacement, and violent persecution, underscoring how racial scapegoating through disinformation can have devastating, tangible impacts.
- **Fueling Polarization**: Disinformation both results from and exacerbates political and social polarization, creating a vicious cycle. Hunter (2023) notes that disinformation spread through social media can weaken democracy by increasing political polarization, encouraging political violence, and undermining trust in democratic institutions and norms. Disinformation, especially when spread in echo chambers, distorts political debates, intensifies ideological polarization, and impacts democratic processes by misinforming the public during elections (Lombana-Bermúdez et al., 2021). This polarization makes it more challenging to reach a consensus on key societal issues, further fragmenting society.
- **Fueling Societal Distrust**: Beyond media distrust, disinformation can fuel broader societal distrust, eroding confidence in political institutions and democratic processes. For example, Pickel et al. (2022) found that disinformation can increase resentment towards minorities, alienate individuals from democracy, and bolster support for authoritarian parties. Similarly, Berlinski et al. (2023) argue that unsubstantiated claims of voter fraud reduce confidence in electoral integrity and contribute to the erosion of democratic norms. This erosion of trust can lead to a decline in civic engagement and a more polarized and fragmented society.

Some of the impacts discussed here—like polarization or distrust in institutions—also showed up earlier in our analysis as enabling factors or contextual vulnerabilities. This overlap highlights the existence of feedback loops, where the effects of disinformation don't just cause harm, but actually reinforce the conditions that made society vulnerable in the first place. This suggests that, as information disorder persists, the information ecosystem, society and democracy, becomes increasingly more susceptible to further disinformation.

**Discussion**

The Framework for Information Disorder presented in this article provides a structured approach to explain the mechanisms and implications of information disorder. By distinguishing between clearly defined stages—enabling vulnerabilities, agents and motivations, tactical strategies, informational impacts, and societal consequences—the framework aims to facilitate a more holistic understanding of the cases contributing to information disorder.

Context emerges as crucial for understanding how information disorder thrives. Vulnerabilities at the sociopolitical and informational levels are context-specific and enable information disorder to thrive. Recognizing why particular narratives resonate—and conversely, how to construct meaningful counter-narratives—depends heavily on these contextual factors. Thus, effective responses can not only rely on addressing the false *content* but need to be based on a thorough understanding of the enabling vulnerabilities that are allowing the narrative to take root, resonate with the audience and spread.

A core contribution of this framework is its emphasis on orchestration, highlighting the intentionality behind disinformation efforts. Understanding the agents involved —their motivations, resources, and strategies—reveals that disinformation is rarely about outright falsehood alone; rather, it blends truths, half-truths, and distortions to strategically undermine trust in information systems. Therefore, the term "information disorder" captures the systemic degradation of the information environment, emphasizing its broader implications for democratic institutions and societal cohesion.

Additionally, our framework underscores the importance of considering the broader informational environment as a whole, moving the focus away from social media only and integrating the cross-cutting role of the media. Of course, digital platforms remain central and have effectively transformed the informational landscape, by enabling faster dissemination, targeted manipulation, and monetization. Still, media entities—traditional and digital—are active stakeholders that must be taken into account.

Finally, it is important to emphasize that information disorder not only directly transforms the information ecosystem but also has broader implications for society and democracy, affecting political dynamics, public policy decisions, and democratic institutions (politics, policy, and polity). We find that information disorder directly impacts politics by influencing electoral outcomes, voter behavior, and political discourse; indirectly but significantly shapes policy by distorting policy-making processes and altering public perceptions on issues like public health or immigration; and undermines polity by eroding institutional trust, intensifying polarization, and weakening societal consensus around democratic norms. This multifaceted impact highlights how information disorder operates systematically, reinforcing conditions that enable its proliferation, and thus creating self-sustaining feedback loops of societal vulnerability.

Limitations of this framework must be acknowledged. First, the complexity of information disorder makes it challenging to capture all possible variations and nuances within a single framework. Disinformation campaigns are highly adaptive and context-specific, and while this framework provides a robust starting point, it may not account for emerging (and ever evolving) tactics or unforeseen contexts. Second, the difficulty in establishing a direct, causal relationship between disinformation and its societal impacts remains. As part of future work, we are interested in applying this framework to concrete cases of information disorder — particularly regarding public health, climate change and gender.

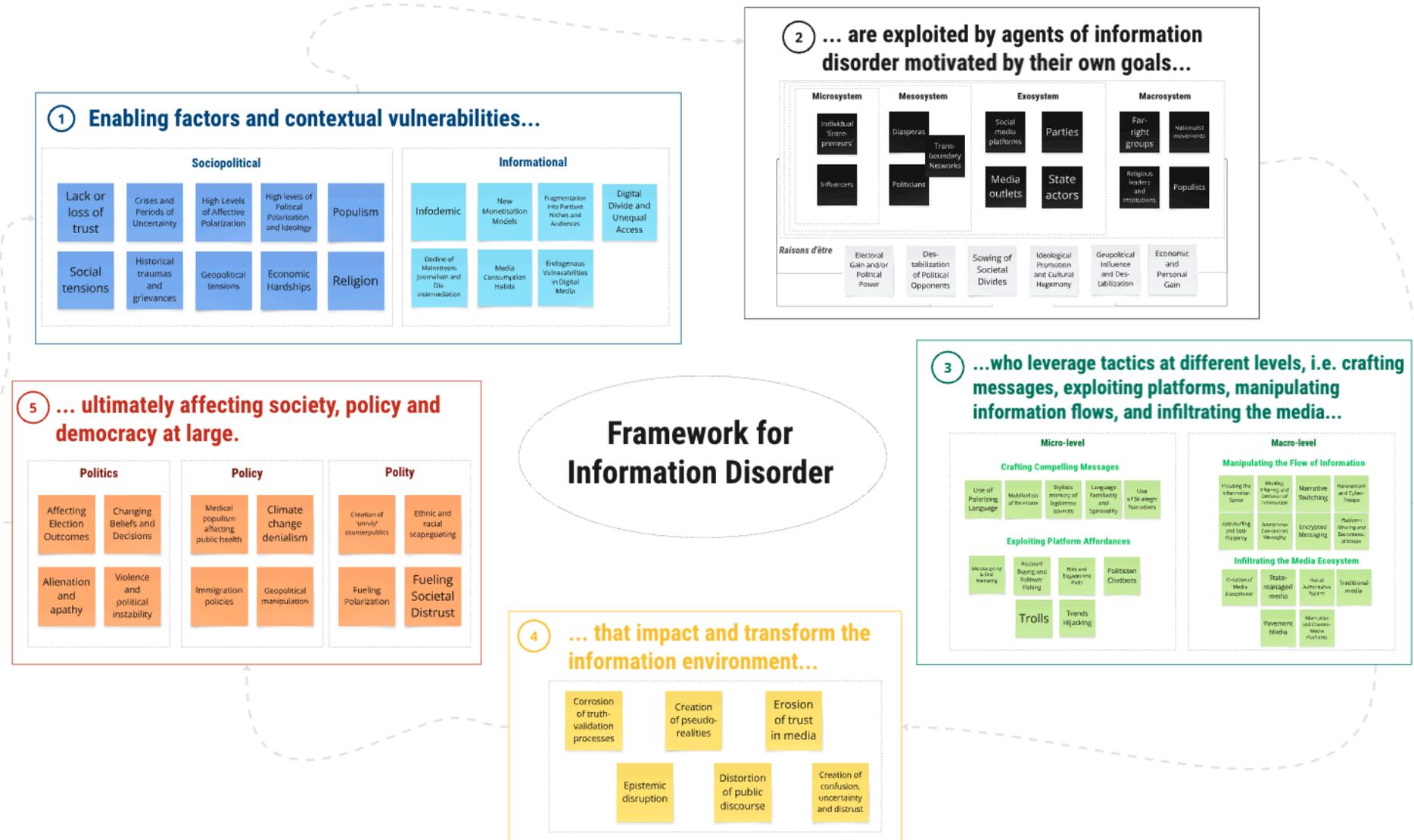

*Figure 7: Full Framework for Information Disorder. Source: Elaborated by the authors.*